\documentclass[twocolumn, 10pt, pra, aps, showpacs, reprint, amsmath, amssymb, superscriptaddress, nofootinbib]{revtex4-1}

\usepackage[utf8]{inputenc}
\usepackage[english]{babel}
 
\usepackage[]{graphicx}

\usepackage{amsfonts}
\usepackage{amsmath, bm}

\begin{document} 
\title{Multimode lasing in wave-chaotic semiconductor microlasers}
\author{Alexander Cerjan}
\affiliation{Department of Physics, The Pennsylvania State University, University Park, PA 16802, USA}
\author{Stefan Bittner}
\author{Marius Constantin}
\affiliation{Department of Applied Physics, Yale University, New Haven, Connecticut 06520, USA}
\author{Mikhail Guy}
\affiliation{\ Yale Center for Research Computing, 
Yale University, New Haven Connecticut 06520}
\author{Yongquan Zeng}
\author{Qi Jie Wang}
\affiliation{Center for OptoElectronics and Biophotonics, School of Electrical and Electronic Engineering and Photonics Institute, Nanyang Technological University, 639798 Singapore}
\author{Hui Cao}
\author{A. Douglas Stone}
\email{Corresponding author. Email: douglas.stone@yale.edu}
\affiliation{Department of Applied Physics, Yale University, New Haven, Connecticut 06520, USA}

\begin{abstract}
We investigate experimentally and theoretically the lasing behavior of dielectric microcavity lasers with chaotic ray dynamics. Experiments show multimode lasing for both D-shaped and stadium-shaped wave-chaotic cavities. Theoretical calculations also find multimode lasing for different shapes, sizes and refractive indices. While there are quantitative differences between the theoretical lasing spectra of the stadium and D-cavity, due to the presence of scarred modes with anomalously high quality factors, these differences decrease as the system size increases, and are also substantially reduced when the effects of surface roughness are taken into account. Lasing spectra calculations are based on Steady-State Ab Initio Laser Theory, and indicate that gain competition is not sufficient to result in single-mode lasing in these systems.
\end{abstract}

\maketitle

\section{Introduction}
\label{introduction}

There has been a great deal of interest in the properties of dielectric microcavity lasers or resonators based on quasi-two-dimensional cavities, for which  different boundary shapes can generate chaotic, mixed or regular ray dynamics \cite{CW_RMP}, with corresponding implications for the resonant wave solutions and lasing modes of such cavities. We will use the term {\it wave-chaotic cavity} to refer to cavity shapes for which the ray dynamics based on specular reflection of rays at the boundary and neglecting refractive escape satisfy standard definitions of chaotic or partially chaotic dynamics \cite{Altmann2013RMP}. The motivation for this work are recent experimental results and theoretical arguments concerning semiconductor microlasers with fully chaotic ray dynamics (i.e., no stable periodic orbits or quasi-periodic Kolmogorov–Arnold–Moser, or KAM orbits), and in this article we will only treat cavity shapes with either fully chaotic ray dynamics or with fully integrable ray dynamics (such as the ellipse). It should be emphasized that we do not concern ourselves with the temporal dynamics, which for semiconductor lasers can be unstable or chaotic for a number of reasons \cite{Ohtsubo2013BOOK}, and in the theoretical work presented below we confine ourselves to the study of single- or multimode steady-state solutions of the laser equations. 

The earliest dielectric cavity lasers were microdisk lasers with whispering gallery type high-$Q$ modes confined by total internal reflection \cite{McCall1992}. Not long after these first microdisk lasers were demonstrated, the idea of  deforming the resonator boundary to non-circular shapes was introduced \cite{Nockel94OptLett, Nockel1997Ray} to explore the implications of fully or partially chaotic ray dynamics on the wave solutions and lasing properties. A major focus of this earlier work was on obtaining directional emission from smoothly deformed cavity shapes, and on using properties of the phase space of the ray dynamics to predict and control the emission patterns \cite{Rex2002PRL, Nockel1996Directional, Wiersig2006Unidirectional}. In addition, many of these early studies considered laser cavities much larger than the wavelength for which one could not resolve individual lasing lines, motivating a statistical treatment via ray escape models. However when wave-chaotic microcavities with sizes of the order of the wavelength were fabricated and studied \cite{SongPRL2010, SongPRA2011}, the lasing spectrum was generally found to be multimode \cite{FangAPL2007, SongPRA2009}. The nature of their lasing modes and their relationship to passive cavity modes and the classical ray dynamics have been studied extensively \cite{CW_RMP}. 

Motivated by wave-chaotic and random microcavity lasers the Steady State Ab Initio Laser Theory (SALT) was developed starting in 2006 \cite{Tureci2006PRA, Tureci2007PRA, Tureci2008Science, Ge2010SALT, esterhazy2014scalable, cerjan2015steady}, an approach which can be employed for arbitrary complex geometries and yields the active mode spectra of microcavity lasers. SALT is based on the Stationary Inversion Approximation (SIA), which becomes reasonable for sufficiently small cavities, as emphasized first by Fu and Haken \cite{HakenFu1988Semiclassical}. The quantitative validity of SALT for microlasers in the appropriate regime has been confirmed by comparison with full time-dependent solutions of the two-level and multi-level semiclassical laser equations \cite{Ge2008Quantitative}. The theory is designed to describe multimode steady-state lasing and takes gain competition and gain saturation into account to all orders, within the SIA. SALT and a further "single-pole approximation", known as SPA-SALT, has been used to study how spatially selective pumping can be used to control the lasing spectra of wave-chaotic, random and circular microlasers \cite{Ge2010SALT}. 

More recently, wave-chaotic cavity lasers or random lasers have been proposed and demonstrated by several of the authors as novel bright sources emitting spatially incoherent light for imaging and microscopy applications \cite{Redding2015PNAS, Redding2012Speckle, Mermillod-Blondin2013OPTLET, Cao2019Review}. For these applications, wave-chaotic GaAs D-cavity lasers with sizes of the order of $10^2-10^3 \mu \rm{m}$ and pulsed electrical pumping were tested and compared to circular disk and Fabry-Perot cavity lasers fabricated in a similar manner \cite{Redding2015PNAS}. Speckle contrast measurements of the D-laser emission indicated $N_M \sim 10^2 - 10^3$ lasing modes with distinct spatial profiles, many orders of magnitude greater than the number of spatial modes found in traditional lasers such as Fabry-Perot broad-area lasers of comparable surface area. The incoherent superposition of so many different spatial modes results in low spatial coherence which suppresses coherent artifacts. It was thus argued that wave-chaotic microlasers were particularly good devices for generating highly-multimode, spatially incoherent lasing emission, since they exhibit many spatially distinct modes with similar $Q$-factors, and their modes have speckle-like intensity distributions filling the entire cavity, utilizing the entire available gain medium. 

Very different lasing behavior was found by Sunada \textit{et al.}\ in other recent experiments with stadium-shaped GaAs microlasers of a somewhat smaller size. In Ref.~\cite{2013SunadaPRA} it was found that the stadium microlasers transitioned from multimode lasing when pumped with very short pulses to single-mode lasing for pump pulses longer than $100~\mu$s. Experiments in the steady-state regime \cite{Sunada2016PRL} with continuous wave (cw) pumping showed {\it single-mode} lasing for the wave-chaotic stadium lasers, whereas multimode lasing was found for elliptical microlasers (which have integrable ray dynamics). The authors argued, based on numerical calculations, that the modes of wave-chaotic resonators overlap so strongly in space that their cross gain-saturation results in single-mode lasing, whereas whispering gallery modes with different radial quantum numbers in the elliptic resonator have small enough overlap to allow multimode lasing. Subsequent theoretical and numerical work indicated that single-mode lasing may be typical for wave-chaotic microlasers in certain parameter regimes \cite{Harayama2017PhotonRes}. These two sets of experiments (Refs.~\cite{2013SunadaPRA, Sunada2016PRL} and Ref.~\cite{Redding2015PNAS}) raise fundamental questions about the lasing dynamics of wave-chaotic semiconductor microlasers, and the aim of this article is to elucidate the influence of different experimental parameters and physical mechanisms on the number of lasing modes given that the microlasers used in the two sets of experiments are very similar, though not identical. In particular, we consider the effects of mode competition, the size, refractive index and surface roughness of the cavities, as well as non-universal features of different wave-chaotic resonator geometries. 

In the first part of the article we present new experimental measurements of the lasing spectra of both D-cavity and stadium-shaped wave-chaotic GaAs microlasers with time resolution in order to address the questions raised above. In the second part of the paper we present the results of SPA-SALT calculations that shed light on the role of mode competition in wave-chaotic cavity lasers. It should be noted that the microlaser cavities used in experiments are too large to be directly simulated either by time domain or frequency domain (e.g., steady-state) approaches. Moreover, SALT loses quantitative validity when the lasing spectra becomes too dense with increasing cavity size. Hence our calculations do not allow quantitative modeling, but provide evidence and physical insight into the role of wave chaos and mode competition in multimode lasing. 

\section{Experimental Results}
\label{experimental}

\begin{figure}
\begin{center}
\includegraphics[width = 8.4 cm]{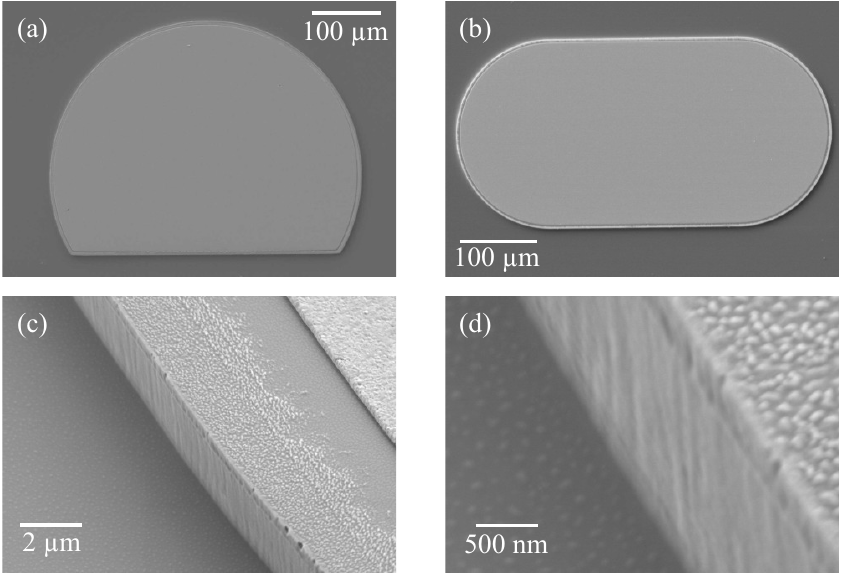}
\end{center}
\caption{Top view SEM images of (a)~a D-cavity with radius $R = 200~\mu$m and (b)~a stadium cavity with $L = 238~\mu$m. (c,~d)~Perspective SEM images of the curved sidewall of a D-cavity, highlighting its verticality and low surface roughness.}
\label{fig:SEM}
\end{figure}

We investigated edge-emitting GaAs quantum well semiconductor lasers fabricated from a commercial epi-wafer (Q-Photonics QEWLD-808) by photolithography and dry etching (see Ref.~\cite{Bittner2018Science} for details of the fabrication process). Microlasers in the shape of a D-cavity, a stadium and an ellipse were created. Scanning-electron microscope (SEM) images of two cavities are shown in Figs.~\ref{fig:SEM}(a, b), respectively. The dry-etching process ensured vertical sidewalls [see Fig.~\ref{fig:SEM}(c)] and a low degree of surface roughness [see Fig.~\ref{fig:SEM}(d)]. 

\begin{figure}
\begin{center}
\includegraphics[width = 8.4 cm]{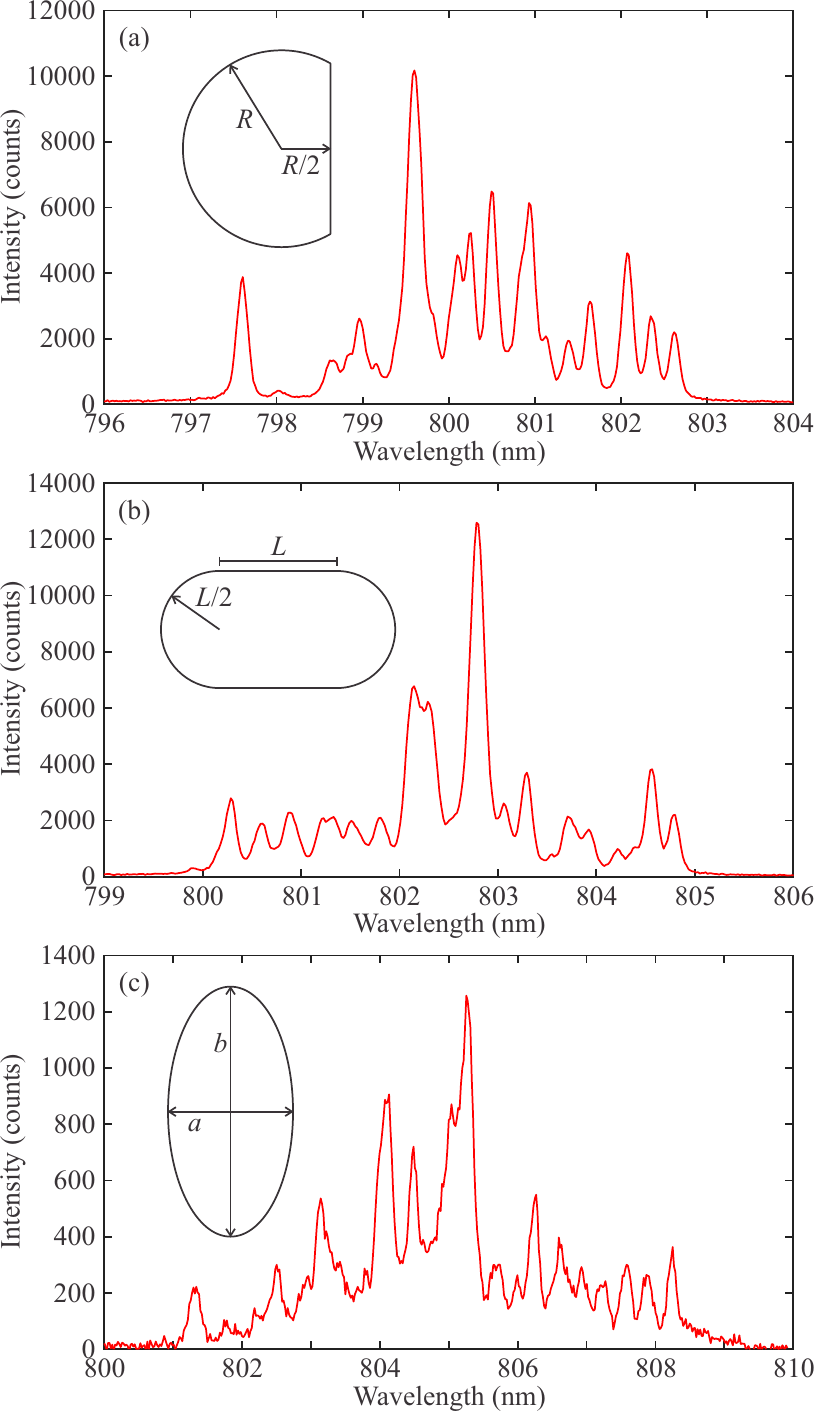}
\end{center}
\caption{Lasing spectra integrated over a $2~\mu$s-long pump pulse with $500$~mA pump current for (a)~a D-cavity with $R = 100~\mu$m, (b)~a stadium with $L = 119~\mu$m, and (c)~an ellipse with $a = 127~\mu$m and $b = 254~\mu$m. The insets illustrate the geometry of the cavities.}
\label{fig:spectra-TI}
\end{figure}

The D-cavity shape is a circle with radius $R$ from which a segment has been cut off $R / 2$ away from the center as shown in the inset of Fig.~\ref{fig:spectra-TI}(a). The stadium cavities considered in this article consist of a square with side length $L$ to which two semicircles with radius $L / 2$ are attached as shown in the inset of Fig.~\ref{fig:spectra-TI}(b). The ray-dynamics of both cavity types is fully chaotic \cite{Bunimovich1979Ergodic, Reichl1999Dshape}. The ellipse cavities we investigated have an aspect ratio of $b / a = 2$ where $a, b$ are the two diameters [see inset of Fig.~\ref{fig:spectra-TI}(c)]. Their ray dynamics is integrable. 

The cavities were mounted on a large Cu block which acted as a passive heat sink and pumped electrically with $2$ to $500~\mu$s-long pulses by a diode driver (DEI Scientific PCX-7401). All experiments were performed at ambient temperature. The laser emission from the cavities was collected by an objective and transmitted to an imaging spectrometer (Acton SP300i) with a multimode fiber bundle. An intensified CCD camera (ICCD, Andor iStar DH312T-18U-73) attached to the spectrometer was used to measure the evolution of the emission spectrum during a pulse with microsecond resolution. The experimental setup and measurement procedures are described in more detail in Ref.~\cite{Bittner2018Science}. 

The lasing spectra integrated over $2~\mu$s-long pulses for a D-cavity, a stadium and an ellipse are shown in Fig.~\ref{fig:spectra-TI}. All three cavities have approximately the same area of $25,300~\mu$m$^2$ and hence the same resonance density. The spectra for $500$~mA pump current show multimode lasing with about $20$ peaks for all three geometries. The emission has transverse electric (TE) polarization with the electric field parallel to the cavity plane. Similar results were obtained for cavities with two times larger linear dimension. The actual number of lasing modes cannot be determined from the spectra as our spectrometer cannot resolve closely-spaced lasing modes due to its finite spectral resolution. Nevertheless, the appearance of multiple peaks in the emission spectrum clearly evidences multimode lasing in D-cavities and stadia of this size. 

\begin{table}
\begin{tabular}{lcc}
\hline \hline
Cavity & $\langle I_{th} \rangle$ (mA) & $\langle j_{th} \rangle$ (A~cm$^{-2}$) \\
\hline
D-cavity, $R = 100~\mu$m & $125 \pm 6$ & $494.6$ \\
Stadium, $L = 119~\mu$m & $100 \pm 2$ & $395.5$ \\
Ellipse, $a = b/2 = 127~\mu$m & $68 \pm 4$ & $269.6$ \\
\hline \hline
\end{tabular}
\caption{Average threshold currents $\langle I_{th} \rangle$ and corresponding threshold current densities $\langle j_{th} \rangle$ for cavities with the same size as those presented in Fig.~\ref{fig:spectra-TI}.}
\label{tab:thresholds}
\end{table}

The threshold currents vary very little between cavities of the same type and size, but depend on the cavity geometry. An overview of the average threshold currents $\langle I_{th} \rangle$ for the three cavity shapes shown in Fig.~\ref{fig:spectra-TI} is given in Table~\ref{tab:thresholds}. The stadium cavities have somewhat lower thresholds than the D-cavities. The ellipse cavities have clearly lower thresholds than the two wave-chaotic cavities. We would, however, expect orders of magnitude lower lasing thresholds for ellipse cavities due to the existence of whispering gallery modes (WGMs) with ultra-high $Q$-factors. We attribute the only moderate difference to the small, but not negligible surface roughness. These results motivate a detailed study how the cavity geometry and surface roughness determine the passive mode quality factors presented further below. 

\begin{figure*}
\begin{center}
\includegraphics[width = 16 cm]{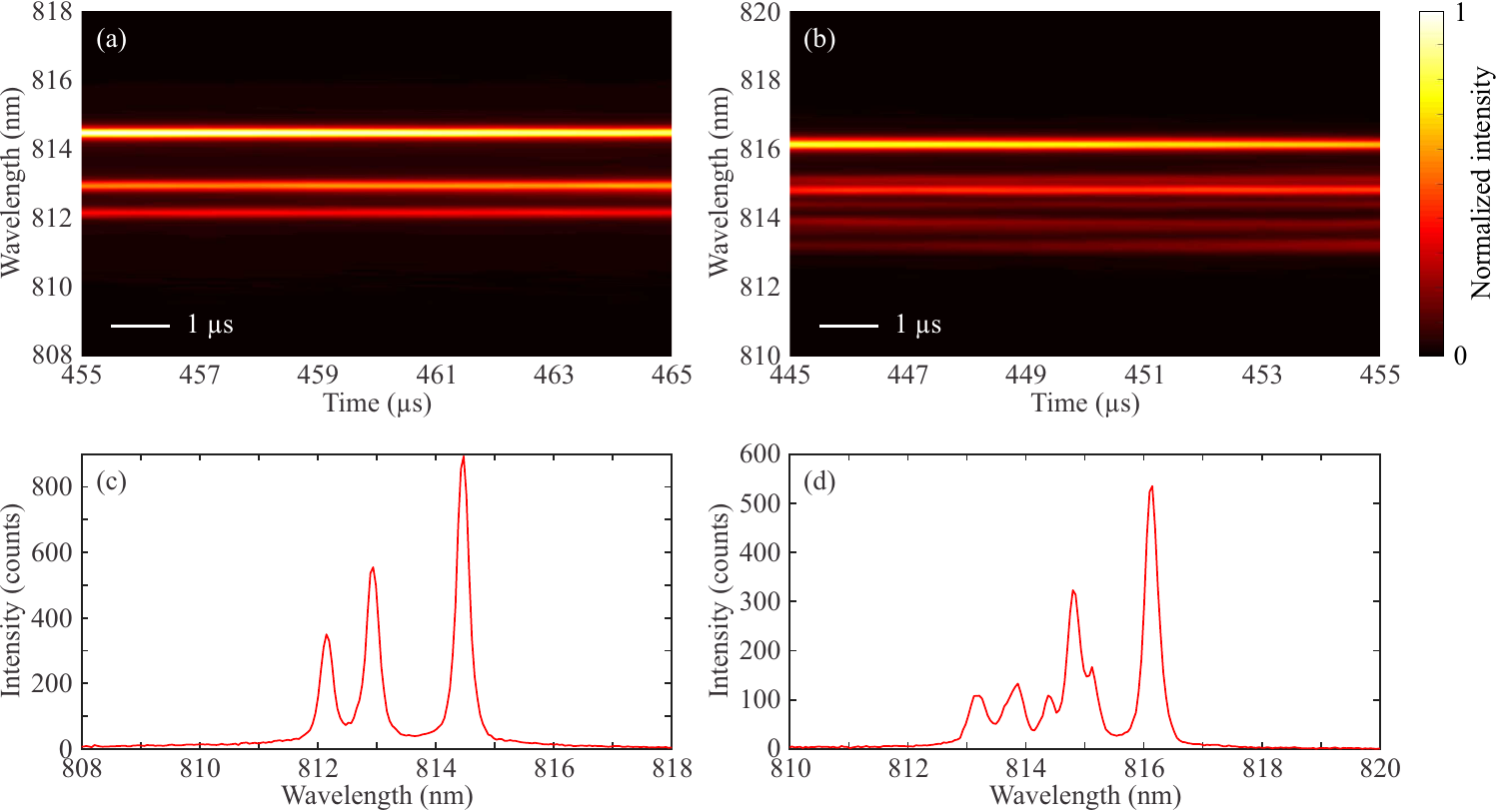}
\end{center}
\caption{Spectrochronogram of (a)~a D-cavity with $R = 200~\mu$m and (b)~a stadium with $L = 238~\mu$m during a $500~\mu$s-long pulse measured with $1~\mu$s time resolution. The emission spectra show no discernible change over the course of $10~\mu$s. The pump current was $800$~mA in both cases. (c)~Emission spectrum of the D-cavity during the time interval $460$--$461~\mu$s and (d) of the stadium during the time interval $450$--$451~\mu$s, showing multiple lasing peaks.}
\label{fig:spectraTR}
\end{figure*}

For a detailed understanding of the multimode lasing dynamics, it is important to take into account thermal effects. The cavities heat up quickly during current injection, which leads to a red-shift of the emission spectrum and changes of the active lasing modes during the pump pulse \cite{Bittner2018Science}. The temperature and hence the emission spectrum gradually stabilizes, however, over the course of longer pulses. A D-cavity with $R = 200~\mu$m and a stadium with $L = 238~\mu$m were pumped with $500~\mu$s-long pulses at $800$~mA pump current. The threshold currents are $I_{th} = 270$~mA ($j_{th} = 267.1$~A~cm$^{-2}$) for the D-cavity and $I_{th} = 230$~mA ($j_{th} = 227.4$~A~cm$^{-2}$) for the stadium, respectively. Both cavities have approximately the same area of $101,100~\mu\mathrm{m}^2$. Excerpts of the spectrochronograms of the D-cavity and the stadium measured with $1~\mu$s time resolution are shown in \mbox{Figs.~\ref{fig:spectraTR}(a, b)}, respectively.  After more than $400~\mu$s, the lasers have stabilized so well that the emission spectrum does not change over the course of $10~\mu$s. It has also been verified that the lasers do not exhibit any fast, nanosecond time-scale dynamics \cite{Bittner2018Science}. Thus, a quasi-steady state of the lasing dynamics is reached. Even though thermal equilibrium is not reached due to the lack of active cooling, the remaining thermally-induced fluctuations of the emission spectrum with a time scale longer than $10~\mu$s are several orders of magnitude slower than the intrinsic dynamical time scales of semiconductor lasers. Hence, a steady-state model such as SALT is appropriate to investigate the interactions of the lasing modes via the active medium in this regime. 

The spectra of both D-cavity and stadium lasers at any given time exhibit multiple lasing peaks as shown in the spectra in Figs.~\ref{fig:spectraTR}(c, d). It should be noted that the actual number of lasing modes is higher than the number of peaks observed in the spectra due to the finite resolution of the imaging spectrometer. Both stadium and D-cavities evidently always exhibit multimode lasing as proven by the presence of several peaks in the spectra. Furthermore, the typical number of lasing peaks is similar for both cavity geometries, and there is no qualitative difference between the lasing spectra of D-cavities and stadia. 

The similarities between D-cavity and stadium lasers are attributed to the common features of wave-chaotic modes in these resonators. However, the lasing modes typically correspond to the highest-$Q$ modes, which may have different characteristics than the majority of the resonance spectrum  in wave-chaotic cavities. Specifically, it is known that wave-chaotic microlasers can exhibit lasing on localized scar modes \cite{Gmachl2002OptLett, Harayama2003SCARS}, which can have higher $Q$-factors than expected from statistical analysis of wave-chaotic lasers \cite{FANG2005PRA, Novaes2012PRE}. Such effects will vary with the cavity shape and size and may be affected significantly by surface roughness. Therefore, a detailed analysis of geometry-specific properties of wave-chaotic microlasers as well as the effect of surface roughness on the lasing dynamics is needed to shed further light on the experimental results. 

\section{Theoretical Study}
\label{theoretical}  

The experimental results presented above are consistent with earlier work by our group on wave-chaotic D-cavity lasers \cite{Redding2015PNAS}, and provide more insight into the time evolution of the lasing spectra. They confirm the qualitatively different behavior of the lasers we have studied from those reported in references \cite{2013SunadaPRA, Sunada2016PRL}, and these different results present a challenge to obtain a consistent understanding of the lasing behavior of wave-chaotic semiconductor lasers. As noted earlier, there is no possibility to simply perform ab initio integration of the laser equations for two-dimensional complex cavities with a size of over $100~\mu$m. The largest cavities to be treated by brute force integration are two orders of magnitude smaller in linear dimensions \cite{Bidegaray2003Time, Huang2006OptExp, Cerjan2015OptExpLASER} and thus have a four order of magnitude smaller resonance density than the experimental microlasers. Our partially analytic approach based on SALT can do somewhat better, providing results for resonators on a $60 \mu \rm{m}$ scale, but with some caveats described below. Thus there is at present no absolutely rigorous computational method \cite{Bohringer2008PQE} to decide the validity of the conjecture in Refs.~\cite{2013SunadaPRA, Sunada2016PRL, Harayama2017PhotonRes} that fully wave-chaotic cavities are intrinsically single-mode lasers in steady-state. 

However, one can use existing methods for such lasers at smaller scale to analyze the physical processes which are known to determine the number of lasing modes. Multimode lasing is known to result from the interplay between gain saturation, which tends to clamp the gain once a single mode starts lasing, and spatial hole burning, which refers to the spatial non-uniformity of gain saturation in standing wave cavities. Near the intensity minima of a lasing mode there is only a small field stimulating emission and the local inversion (gain) is not saturated, and therefore continues to increase with increasing pumping in such regions. This allows new, lower-$Q$ modes with different intensity minima or, more generally, weaker spatial overlap to reach threshold and begin lasing. For example, in conventional stable resonators with Gaussian modes, higher order transverse modes have significant intensity away from the optical axis of the resonator, and can thus exploit distinct regions of the gain medium to lase in addition to the fundamental transverse mode. 

In wave-chaotic cavities, resonances tend to fill the entire resonator with a speckle-like intensity distribution. In Ref.~\cite{Sunada2016PRL} it was proposed that significant spatial overlap of the intensity distributions was the origin of single-mode lasing: wave-chaotic modes always overlap so strongly that a single mode clamps the gain in spite of spatial fluctuations of its intensity distribution. The spatially extended nature of the intensity distributions was argued to be the relevant difference between a wave-chaotic shape, such as the stadium, and a non-wave-chaotic shape with similar aspect ratio, such as the ellipse, which showed multimode lasing. This conjecture apparently contradicts the numerical results in Ref.~\cite{Redding2015PNAS} for D-cavity lasers, which found eight modes lased in a D-cavity with area $\approx 63~\mu \mathrm{m}^2$. Here we extend the numerical studies of Ref.~\cite{Redding2015PNAS} to study in more detail the conjecture that strong gain clamping and cross-saturation lead to single-mode lasing in wave-chaotic microlasers.  

The method we use, Steady-State Ab-Initio Laser Theory (SALT), is an approach specifically developed to study microcavity lasers with complex geometries, and was used in earlier works on microcavity lasers \cite{Tureci2006PRA, Tureci2007PRA, Tureci2008Science, Ge2008Quantitative, Ge2010SALT, Cerjan2012NLEVEL}. Not only was SALT developed to treat complex 2D (and in principle 3D) cavities, it was also designed to deal with multimode lasing and spatial hole burning quantitatively. SALT is a semiclassical theory and does not include quantum fluctuation effects, which are not relevant to this study. The version of SALT most relevant to the current work, and its limitations, are described in detail in \cite{Ge2010SALT}. 

\subsection{Review of Steady-State Ab-Initio Laser Theory}
\label{subsection000}
The SALT equations for steady-state multimode lasing are derived from the semiclassical laser equations by neglecting time-dependent non-linear terms in the equations, which drive oscillations in the inversion. They are a set of non-linear wave equations to be solved self-consistently with purely outgoing boundary conditions, and have the form
\begin{equation}
\begin{aligned}
 & \bigg[ \bigg(\nabla \times \nabla \times \bigg) - \bigg( \epsilon_c (\textbf{x}) + \frac{\gamma_{\perp} D_0(\textbf{x})}{\omega_{\mu} - \omega_a + i \gamma_{\perp}}  \\
 & 
 \times \frac{1}{1 + \sum_{\nu}^{N_L} \Gamma_{\nu} |\Psi_{\nu}|^2} \bigg) \frac{\omega_{\mu}^2}{c^2} \bigg] \Psi_{\mu} (\textbf{x}) = 0, \qquad \textbf{x} \in C \, .
\end{aligned}
\end{equation}
They determine the number of lasing modes, their optical field distributions, $\Psi_{\mu}(\textbf{x})$, and  lasing frequencies, $\omega_{\mu}$. The approximation used to derive the SALT equations is that the inversion density (which only appears implicitly in the SALT equations) is time-independent. The inputs to the SALT equations are the dielectric function of the passive cavity, $\epsilon_c(\textbf{x})$, the dephasing rate of the polarization, $\gamma_\perp$, the atomic transition frequency, $\omega_a$, and the external pump profile, $D_0 (\textbf{x})$. For two-level atoms the gain curve is Lorentzian and centered at $\omega_a$ with width $\gamma_\perp$. The Stationary Inversion Approximation (SIA) requires that $\gamma_\perp$ and the typical frequency spacing between lasing modes, $\Delta \omega$, are much larger than the population relaxation rate, $\gamma_\parallel$. The latter condition becomes harder to meet in the highly-multimode regime and for larger laser cavities. In the regime where the SIA holds, excellent agreement is found between SALT and full integration of the semiclassical laser equations \cite{Ge2008Quantitative}. 

Here we have written the SALT equations as scalar equations since we consider modes with transverse magnetic (TM) polarization, for which $\Psi_\nu$ corresponds to the $z$-component of the electric field. (We have calculated Q value distributions for the (TE) modes and find qualitatively similar behavior to that described below). The dielectric function is $1$ outside of the cavity region $C$. The number of non-trivial purely outgoing solutions increases by one at each lasing threshold as the pump $D_0$ is increased starting from zero. The non-linear denominator represents the saturable gain susceptibility, and enforces self-saturation and cross-saturation of the gain in a spatially-varying manner, which takes into account spatial hole-burning exactly. The Lorentzian gain factor of mode $\nu$ is $\Gamma_{\nu} = \gamma_{\perp}^2 / [(\omega_{\nu} - \omega_a)^2 +\gamma_{\perp}^2]$. $\Psi$ and $D_0$  are written in dimensionless form in terms of the natural units of the electric field, $e_c = \hbar \sqrt{\gamma_{\parallel} \gamma_{\perp}} / (2g)$, and the inversion, $d_c = \hbar \gamma_{\perp} / (4 \pi g^2)$ (g is the dipole matrix element of the transition). 

Two main types of algorithms have been developed to solve the SALT equations \cite{Tureci2006PRA, Ge2010SALT,Esterhazy2014PRA}. The first approach, which is the basis for the algorithm used in this work, expands the solution of the SALT equations in a complete set of biorthogonal outgoing wave functions at a given frequency, known as the Threshold Constant Flux (TCF) states. One of these TCF functions is the exact solution of the semiclassical equations at the lasing threshold, denoted as the threshold lasing mode (TLM); the others take into account the change in the spatial pattern of the mode and its non-linear frequency shift above threshold. While this solution method is relatively efficient compared to FDTD \cite{Cerjan2012NLEVEL}, it is still quite computationally expensive when applied to 2D wave-chaotic cavities, so we use two further approximations to SALT which enable us to treat larger laser cavities. 

\subsection{Single Pole Approximation - SPA-SALT}
\label{subsection00}

The first one, the ``Single Pole Approximation" (SPA-SALT) \cite{Ge2010SALT}, assumes that the field distribution and frequency of each lasing mode are fixed to their values at threshold, as given by the TLMs. Hence as the pump is increased only the overall amplitudes of the modes change and need to be determined. Moreover, since SALT neglects the beating of lasing modes, the phases of the modes do not enter the solution, and one can reduce the full SALT equations to the following equations for the intensities, $I_\mu$, of each mode as a function of the pump,
\begin{equation}
\frac{D_0}{D_0^{\mu}} - 1 = \sum_{\nu} \Gamma_{\nu} \chi_{\mu \nu } I_{\nu},
\end{equation}
\begin{equation}
\chi_{\mu \nu } = \int d^2 r \Psi_{\mu}^2 |\Psi_{\nu}|^2.
\end{equation}
Here, the $D_0^{\mu}$ are the non-interacting thresholds of the Threshold Lasing Modes (TLM), which reflect their $Q$-factors and their proximity to the gain center, but neglect effects of gain competition. The coefficients $\chi_{\mu \nu}$ represent modal overlaps. In general they have a small imaginary part for high-$Q$ modes, which we will neglect by using only the real part, $\chi_{\mu \nu} \approx \mathrm{Re}[\chi_{\mu \nu}]$. We rewrite these equations as
\begin{equation}
\frac{D_0}{D_0^{\mu}} - 1 = \sum_{\nu} A_{\mu \nu} I_{\nu},
\label{eq:DAI}
\end{equation}
\begin{equation}
A_{\mu \nu} =  \Gamma_{\nu} \chi_{\mu \nu} \, .
\label{eq:MATRIX}
\end{equation}
The matrix $A_{\mu \nu}$ represents the cross-saturation interaction of all pairs of lasing modes for a given value of the pump power $D_0$. Equation~(\ref{eq:DAI}) is nominally linear, however the set of lasing modes to include at each pump value is not known, and is determined implicitly by the non-linear interactions contained in the $A_{\mu \nu}$. The threshold of the first mode to turn on is given directly by the TLM calculation, but subsequent mode thresholds are determined by the constraint that all intensities fulfill $I_\nu \geq 0$. Hence we must search at each pump value for the largest set of modes which yields positive semi-definite values for the $I_\nu$, and only then invert Eq.~(\ref{eq:DAI}) using the appropriate matrix $A_{\mu \nu}$. Between modal thresholds, when the matrix $A_{\mu \nu}$ is fixed, the non-zero $I_\nu$ vary linearly with pump and are given by
\begin{equation}
I_{\mu} = c_{\mu} D_0 - b_{\mu},
\end{equation}
\begin{equation}
c_{\mu} = \sum_{\nu = 1}^N \frac{(A^{-1})_{\mu \nu}}{D_0^{\nu}},
\end{equation}
\begin{equation}
b_{\mu} = \sum_{\nu=1}^{N} (A^{-1})_{\mu \nu} \, .
\end{equation}
The solution over the full pump range of interest has a kink at each of the thresholds $D_{0, int}^{\mu}$, where the subscript $int$ denotes the threshold in the presence of modal interactions. These interacting thresholds are given by \cite{Ge2010SALT}
\begin{equation}
D_{0, int}^{N} = D_0^{N} \left[ 1 + \sum_{\nu = 1}^{N} A_{N \nu} (c_{\nu} D_{0, int}^{N} - b_{\nu}) \right] \, .
\end{equation}
This equation follows from the condition that as the $N^{th}$ mode turns on its intensity passes through zero from negative values; the matrix $A$ and the constants $b_\nu, c_\nu$ then change appropriately above this pump threshold. This approach has been validated by comparing numerical results between the SPA-SALT and full SALT methods~\cite{Ge2010SALT}. 

\subsection{Resonance SPA-SALT}
\label{subsection0}

The necessary first stage of a SPA-SALT calculation is the calculation of the TLMs and their non-interacting thresholds and frequencies. This can be done by tracking cavity resonances and quasi-modes as the gain increases until they reach the real axis \cite{Ge2010SALT,esterhazy2014scalable, Cerjan2012NLEVEL}. To avoid this step, recently Cerjan \textit{et al}.\ \cite{Cerjan2016Controlling} proposed an analytic approximation for the evolution of poles as the pump increases, so that standard codes for calculating passive cavity modes such as COMSOL can be used. For high-$Q$ modes, the lasing modes will differ little from the passive cavity modes (within the cavity), and thus the passive cavity modes can replace the TLMs in SPA-SALT. We call the resulting method resonance SPA-SALT; a working code for this method is available for download \cite{Cerjan2016ControllingCODE}. This approach was used in our earlier work on D-cavity lasers and was compared with results from full SALT, finding reasonable agreement. We will use this method in the current work to allow us to consider even larger cavities and explore more of the parameter space of interest. Resonance SPA-SALT expresses the non-interacting thresholds needed for SPA-SALT in terms of the complex frequencies of the passive cavity resonances
\begin{equation}
D_{0}^{\mu} =  \Big|  \Big( \frac{\mathrm{Re}[\omega_{\mu}] - \omega_a + i \gamma_{\perp} }{\gamma_{\perp}} \Big) \Big( \frac{ \omega_{\mu}^2 - \mathrm{Re}[\omega_{\mu}]^2}{\mathrm{Re}[\omega_{\mu}]^2}. \Big) \Big | \, .
\label{Rspasalt}
\end{equation}
The lasing frequencies $\omega_{\mu}$ are approximated by the real part of the passive cavity resonance frequencies, and lasing modes within the cavity by the passive cavity mode field distributions. 

\section{RESULTS}
\label{results} 

As noted, the number of lasing modes will be determined ultimately by the passive cavity mode $Q$-factors, the width, $\gamma_\perp$, and center, $\omega_a$, of the gain curve, and by the non-linear interactions between modes due to gain competition/saturation. The $Q$-factor distribution depends only on the passive cavity geometry and refractive index; examples are shown in Fig.~\ref{fig:Q_DIST}.  For resonance SPA-SALT the non-interacting thresholds follow immediately from Eq.~(\ref{Rspasalt}). The effects of mode competition are assessed by analysis of the SPA-SALT lasing equations and their predictions. We will explore both aspects in the following section. 

\subsection{Lasing in Wave-Chaotic Cavities and $Q$-Factor Distributions}
\label{subsection1}

\begin{figure*}[tb]
\begin{center}
\includegraphics[width = 17.5cm]{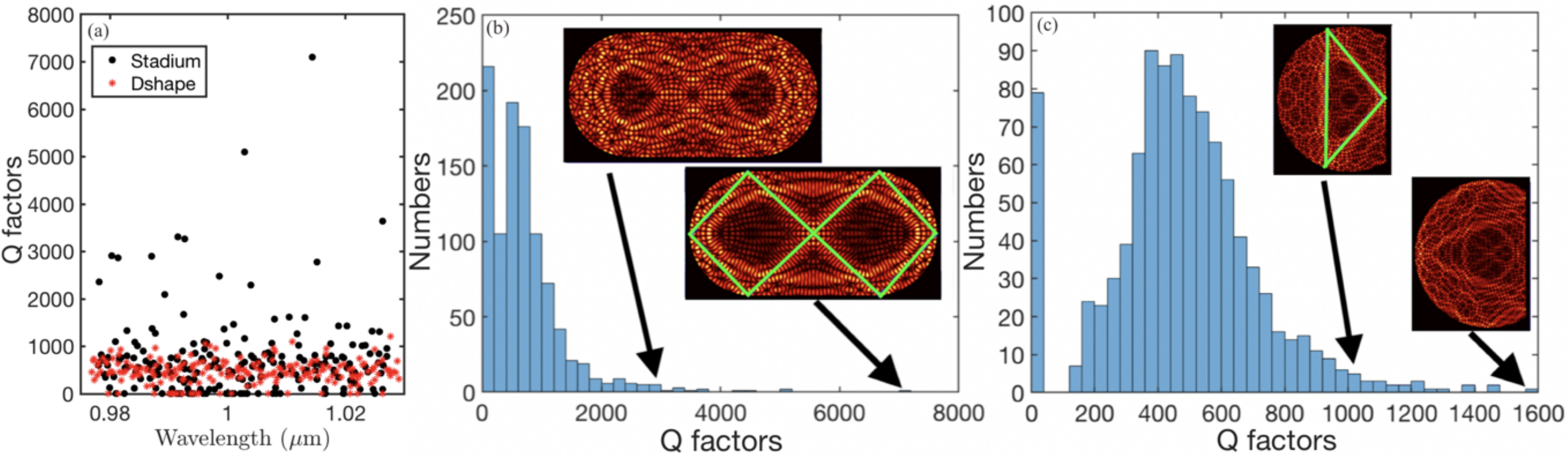}
\end{center}
\caption{(a)~Calculated spectrum ($Q$-factors and wavelengths of passive cavity modes) for the stadium (black circles) and D-cavity (red stars) resonators. The spectra contain $200$ resonances each in a wavelength window centered at $\lambda = 1~\mu$m. The stadium data shows many resonances with higher $Q$-factors than the typical resonances, whereas the D-cavity data does not have any significant outliers. (b)~Distribution of the $Q$-factors of $1000$ resonances for the stadium resonator and the (c)~D-cavity resonator. The insets show the spatial field distributions of the resonances that support the first two lasing modes. The green lines highlight the scarring by the Double Diamond Unstable Periodic Orbit and the Triangle Unstable Periodic Orbit two of the modes.} 
\label{fig:Q_DIST}
\end{figure*}

We begin by examining the $Q$-factor distributions of two resonator geometries, the stadium and the D-cavity, that each have completely chaotic ray dynamics \cite{Bunimovich1979Ergodic}. The stadium consists of a rectangle capped by two semicircles on two opposite sides and is shown in the inset of Fig.~\ref{fig:spectra-TI}(b). Its aspect ratio is defined as the ratio between the length $L$ of the rectangle and the radius $R$ of the semicircles, $\rho_S = \frac{L}{R}$. Our simulation results are for the aspect ratio $\rho_S = 2$, for which the ray dynamics shows the highest degree of chaos, i.e., the highest Lyapunov exponents \cite{Bunimovich1979Ergodic}. The D-cavity, also known as a cut circle, consists of a circle with a part cut off along a single chord as shown in the inset of Fig.~\ref{fig:spectra-TI}(a). We define the cut parameter as is the ratio of the distance from the center of the circle to the chord, $d$, and the radius of the circle, $R$, as $\rho_D = \frac{d}{R}$. In our simulations we use $\rho_D = 0.5$, which also maximizes the degree of chaos of its ray dynamics \cite{Reichl1999Dshape}. 
 
The simulation results for a stadium with $2L = 10~\mu$m and a D-cavity with $R = 4.2~\mu$m, which have approximately the same area, and refractive index $n = 3.5$ are shown in Fig.~\ref{fig:Q_DIST}. Note that in later sections we show results for larger stadium cavities, with a long axis as large as $2L = 60~\mu$m. Figure~\ref{fig:Q_DIST}(a) shows the $Q$-factors of the resonances as a function of the wavelength $\lambda$ in a wavelength window containing $1000$ modes for each resonator geometry. The stadium resonances are indicated by black circles whereas the D-cavity resonances are indicated by red stars. The center wavelength of the window is chosen to be  $1~\mu$m, which also corresponds to the center of the gain curve for the lasing calculations below, $\lambda_a = 2\pi c/\omega_a$. The width of the gain curve used in these simulations is $\gamma_\perp = 50$~nm. This is considerably smaller than wavelength window in which we compute the resonances, $\Delta \lambda_{FW} = 280$~nm.  Hence the simulation results include all the high-$Q$ resonances relevant to lasing. 

Figure~\ref{fig:Q_DIST}(b) shows the distribution of the $Q$-factors of the stadium. The bulk of the distribution shows resonances with $Q$-factors less than $2000$. However, there are many outliers with higher $Q$-factors, and the inset shows the spatial patterns of the first two lasing modes. The first lasing mode is based on the highest-$Q$ resonance ($Q = 7096$), which is localized on an Unstable Periodic Orbit known as the double diamond orbit, a phenomenon known as scarring \cite{Heller1984PRL}. An important aspect is that all the reflections of the double diamond orbit have incidence angles equal to $45^\circ$ and are thus contained by Total Internal Reflection (TIR) for $n = 3.5$. The second lasing mode is based on a resonance with $Q= 3263$ and has a more uniformly distributed spatial pattern. It does not have the second-highest $Q$-factor (or second-lowest non-interacting lasing threshold) as naively expected, so its order in the lasing turn-on sequence is due to its weaker competition with the first lasing mode compared to other potential lasing modes. Figure~\ref{fig:Q_DIST}(c) shows the distribution of the $Q$-factors for the D-cavity, which does not have any significant high-$Q$ outliers. The inset shows the spatial patterns of the first two lasing modes. The first lasing mode is based on the highest-$Q$ resonance ($Q = 1567$), which does not show a strong localization pattern. However, the second lasing mode is based on a resonance with much lower $Q$-factor, $Q = 1100$, which is moderately scarred by the triangle orbit. The incidence angles of the three reflections of the triangle orbit are $47.0^\circ$ and $21.5^\circ$, respectively, and thus this orbit is also contained by TIR for $n = 3.5$. 

\begin{figure*}[tb]
\begin{center}
\includegraphics[width = 8.2cm]{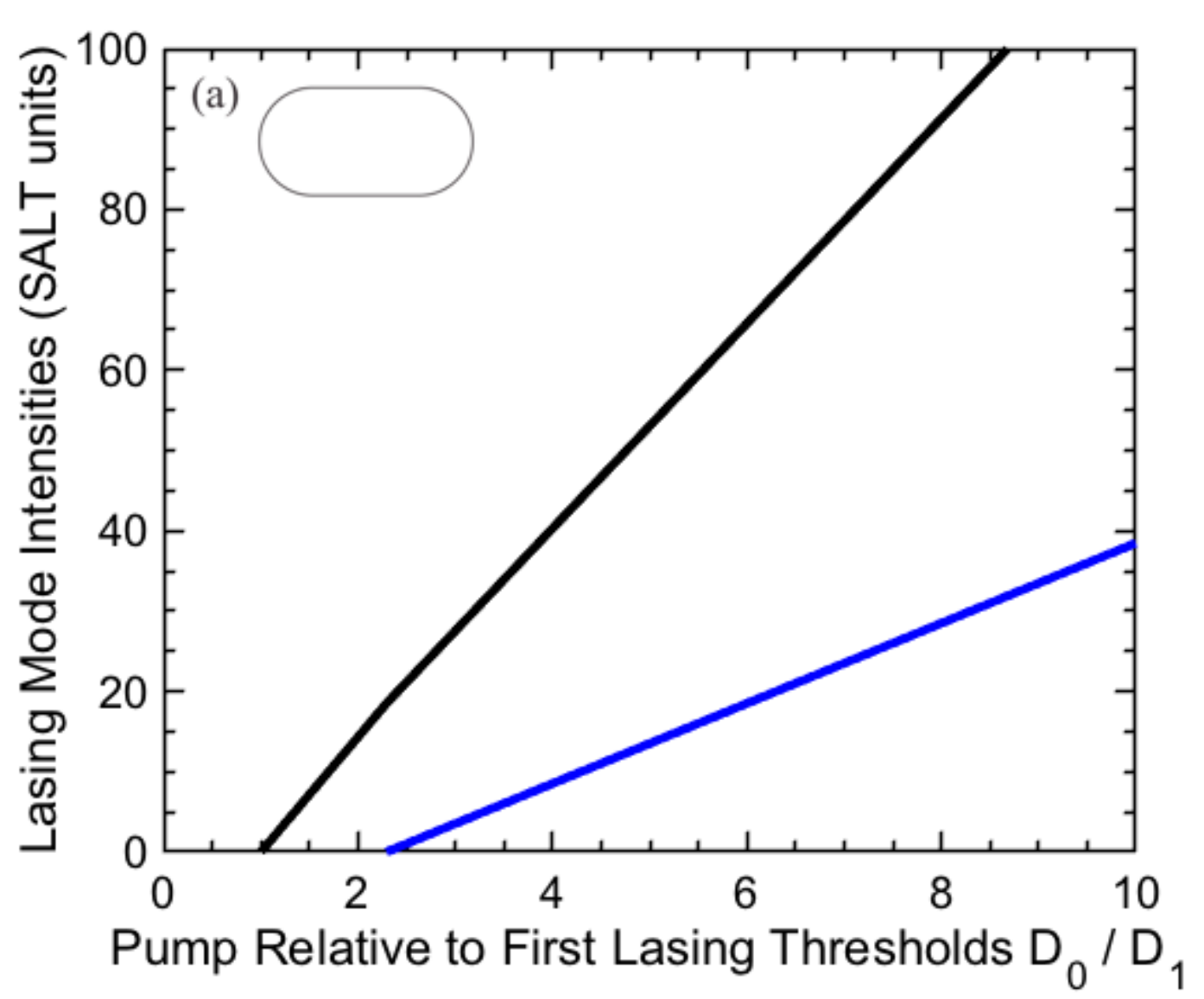}
\includegraphics[width = 8.2cm]{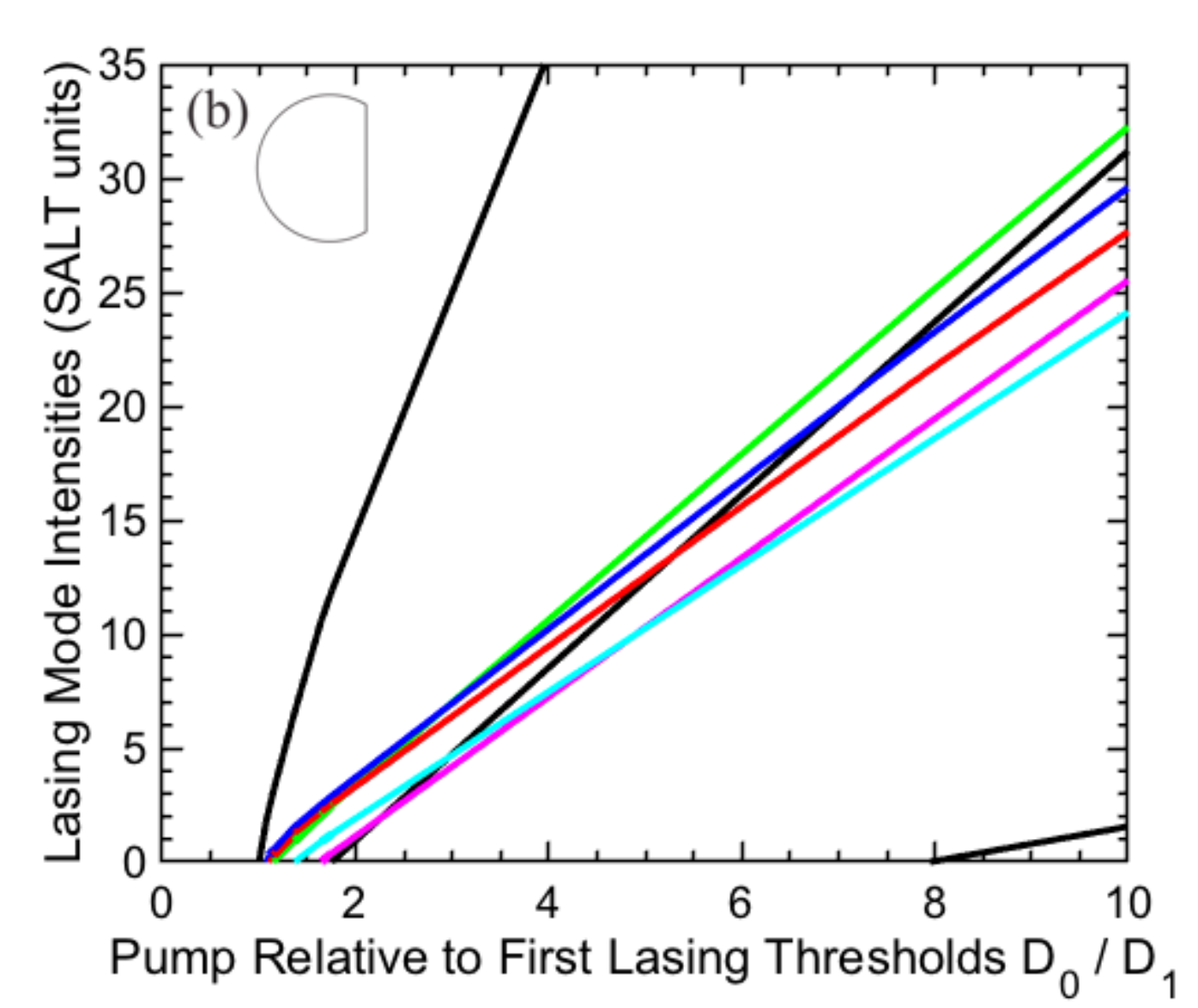}
\end{center}
\caption{(a)~Resonance SPA-SALT results for the lasing intensities of the stadium resonator as a function of the pump strength. There are two lasing modes that turn on within a factor of $10$ of the first lasing threshold. The lasing interactions prevent more modes from turning on in this interval. (b)~Resonance SPA-SALT results for the lasing intensities of the D-cavity resonator as a function of the pump strength. There are eight lasing modes that turn on within a factor of $10$ of the first lasing threshold. The lasing interactions are not strong enough to prevent the first seven lasing modes from turning on within a factor of $2.0$ of the first lasing threshold. The gain spectrum width used in the simulations is $50$~nm.}
\label{fig:PI}
\end{figure*}

Since gain competition does not influence the threshold of the first lasing mode, the ratio of $4.5$ between the highest-$Q$ resonances of the stadium and the D-cavity should lead to a substantially lower lasing threshold for the stadium compared to the D-cavity. Taking into account the wavelength of the highest-$Q$ modes with respect to the gain center yields an about $4.6$ times lower threshold for the stadium compared to the D-cavity. However, experimentally the stadia had an only $1.25$ times lower lasing threshold (see Table ~\ref{tab:thresholds}). The similarity of the measured thresholds for stadia and D-cavities could, however, be explained by the fact that the high-$Q$ scarred modes in the stadium do not exist in much larger cavities or when surface roughness is added as shown further below. 

The $Q$-factor distributions shown here are representative of our results using different wavelength windows and central wavelengths.  Moreover our wavelength window is chosen large enough to contain typical high-$Q$ resonances. For example, the free spectral range (FSR) of modes localized on the shortest periodic orbits (with length $\ell$) in each shape is $\lambda_{FSR} = \frac{\lambda^2}{2  n  \ell} = 28.5$~nm for the stadium and $\lambda_{FSR} = 22.7$~nm for the D-cavity. The high-$Q$ scarred modes correspond to even longer orbits and shorter FSRs, so they are always contained for different central wavelengths $\lambda_a$. 

\subsection{Lasing and Modal Interactions}
\label{subsection2}

\begin{table}[tb]
\begin{tabular}{lccccc}
\hline \hline
Resonator &  \quad n  \quad &  \quad Lasing Modes  \quad &  \quad Gain Clamping Limit \quad \\
\hline
Stadium     & \quad $ 3.5 $& $ 2 $ & $ 6 $  \\   
D-cavity      & \quad $ 3.5 $& $ 8 $  & $ 8 $ \\    
Stadium     & \quad $ 3.0 $& $ 8 $ & $ 9 $ \\   
D-cavity      & \quad $ 3.0 $& $ 14 $ & $ 14 $ \\ 
Stadium     & \quad $ 2.5 $& $ 5 $  & $ 5 $ \\  
D-cavity      & \quad $ 2.5 $& $ 4 $ & $ 6 $ \\  
\hline \hline
\end{tabular}
\caption{Number of lasing modes within a factor of $10$ of the first lasing threshold and number of lasing modes when gain clamping sets in for cavities with different refractive indices.}
\label{tab:index}
\end{table}

Performing SPA-SALT calculations with the resonance data shown in Fig.~\ref{fig:Q_DIST}, we obtain the thresholds, number of lasing modes and mode intensities as a function of pump for a stadium and a D-cavity laser with the same area. Consistent with our expectations and the results of Ref.~\cite{Redding2015PNAS}, both the stadium and the D-cavity lasers exhibit multimode lasing, even with this small size and the full inclusion of gain competition through cross-saturation in the calculations. However, the rather different $Q$-factor distributions of the two wave-chaotic cavities lead to quantitatively different behavior as shown in Fig.~\ref{fig:PI}. For the stadium resonator, only two modes start lasing within a factor of $10$ of the first lasing threshold, whereas for the D-cavity resonator there are eight lasing modes within a factor of $10$ of its first lasing threshold. As already noted, the lasing threshold for the stadium of this size, shape and index is $\sim 4.6$ times lower than for the D-cavity laser, but here we are comparing the number of lasing modes within the same relative range of pump for the two shapes. As we increase the normalized pump strength beyond $10$ we find that a few more modes turn on for the stadium, reaching six modes. But beyond a certain pump strength no additional lasing modes turn on; this phenomenon is known as ``gain clamping'' and has already been reported before in wave-chaotic resonators \cite{Tureci2008Science, Ge2010SALT}.  

Thus our results show that cross-saturation strongly limits the number of lasing modes in wave-chaotic lasers, but does not lead to single-mode lasing at high relative pump values in any case we have studied. If one mode turns on very early because of its anomalously high $Q$-factor, then it is able to saturate the gain substantially before other modes are close to threshold, leading to fewer modes lasing, as in the stadium. For the refractive index $n=3.5$ considered so far the stadium has such outlier modes and the D-cavity does not, leading the latter to have more lasing modes. However this effect depends on the refractive index, and Table~\ref{tab:index} shows that for $n=2.5$ the stadium has more lasing modes within the initial factor of $10$ of relative pump values. The different numbers of lasing modes that we find here for stadium and D-cavity lasers result from non-universal effects of scar modes on short periodic orbits with anomalously high $Q$-factors in the tails of the $Q$-factor distributions. Such effects cannot be described by statistical theories \cite{Keating2008PRA, Schomerus2009}. In the following we explain the different number of lasing modes for stadium and D-cavity lasers by detailed analyses of the cross-gain saturation. Furthermore, we show that the non-universal effects due to high-$Q$ scar modes become weaker for larger cavities and when surface roughness is added. 

\begin{figure*}[!htbp]
\begin{center}
\includegraphics[width = 8.5 cm]{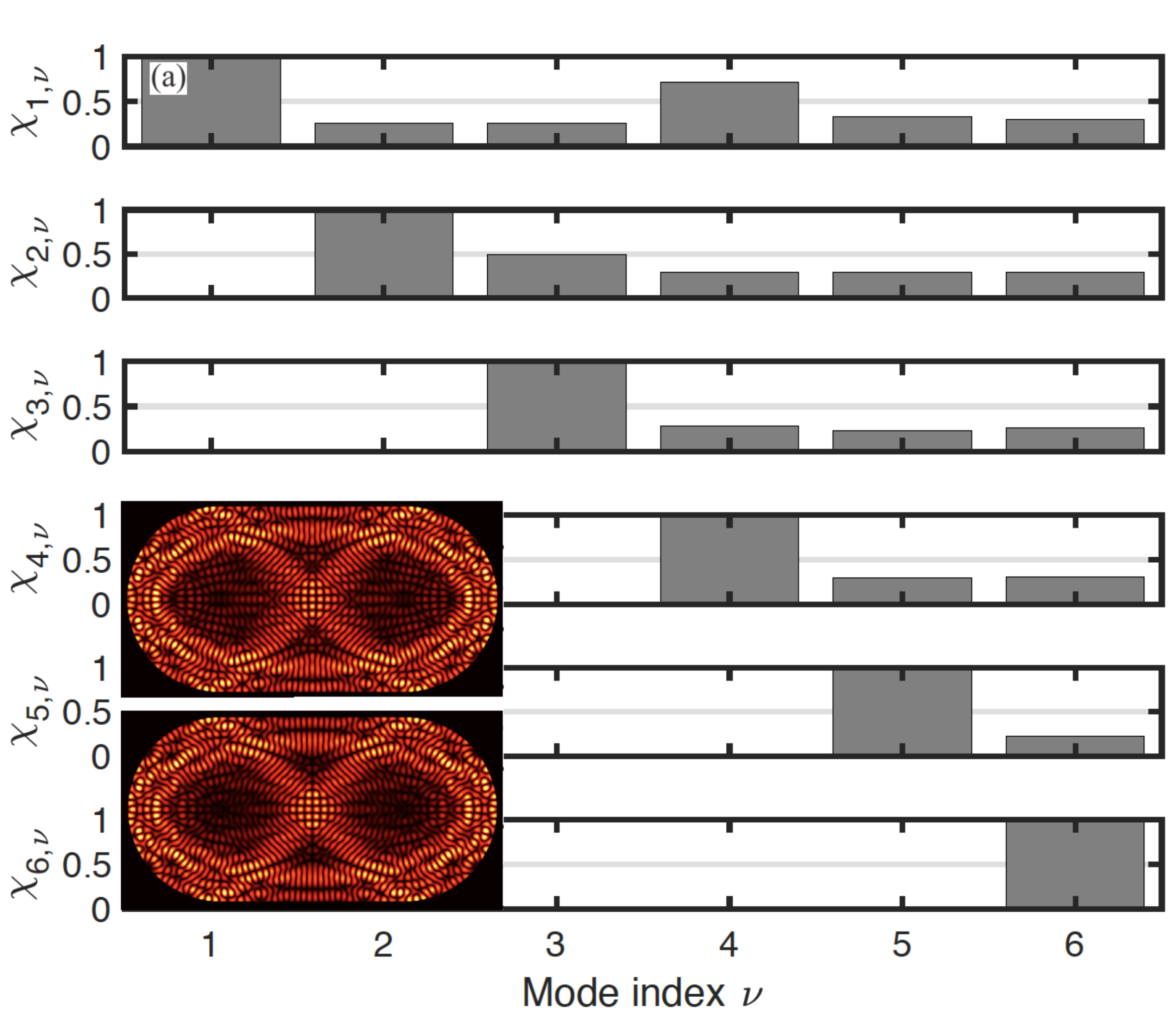}
\includegraphics[width = 8.5 cm]{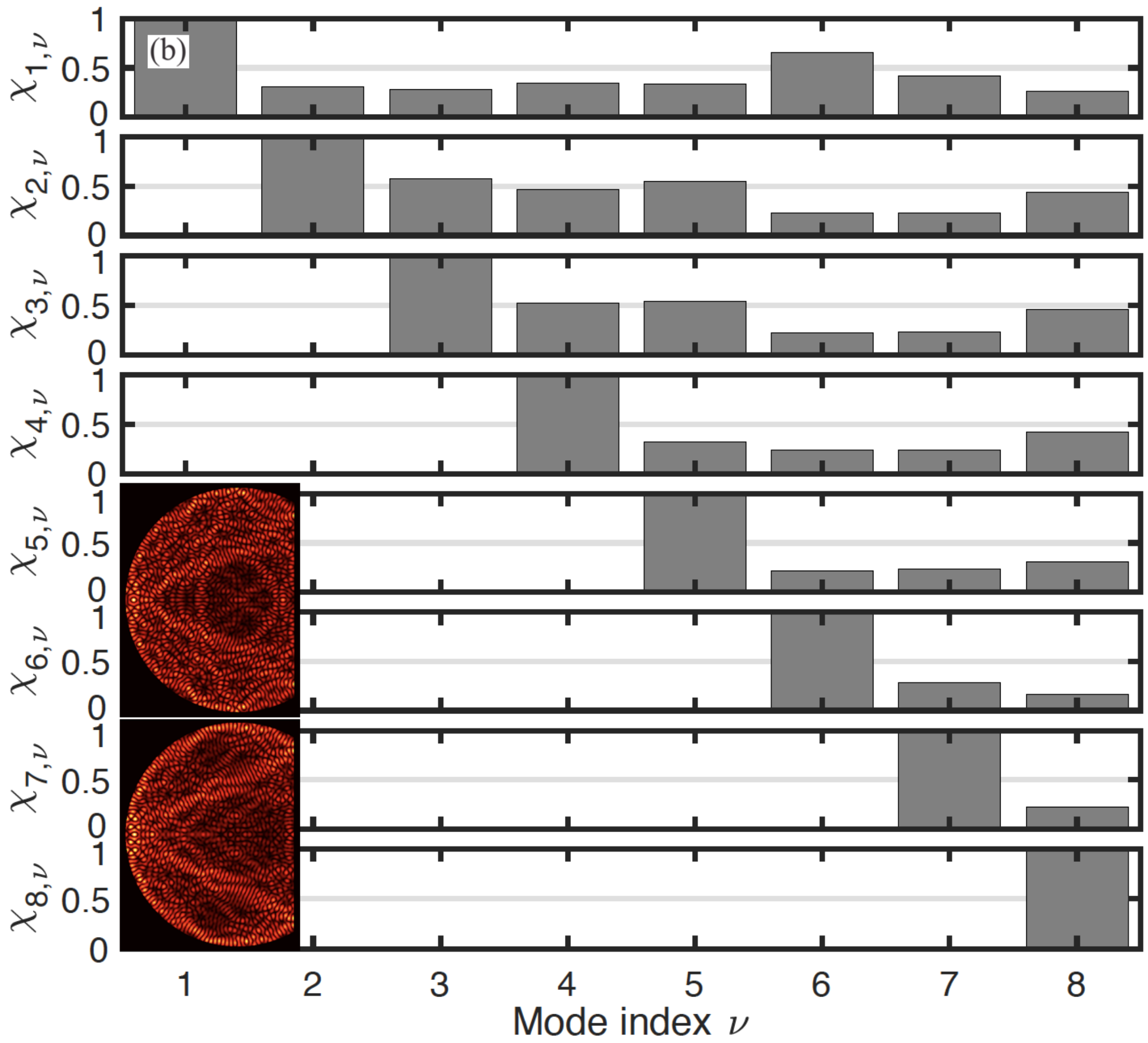}
\end{center}
\caption{Normalized SALT interaction coefficients [Eq.~\ref{eq:chi}] for (a)~stadium and (b)~D-cavity. (a)~For the stadium, the interaction between modes $\#1$ and $\#4$ is higher than the interaction between modes $\#1$ and the other lasing modes, $\#2,3,5,6$, due to their strong spatial overlap (as shown in the inset, top mode $\#1$, bottom mode $\#4$). (b)~For the D-cavity, the interactions are overall more uniform than for the stadium. Modes which have similar spatial patterns form two groups: modes $\#1, 6, 7$ and modes $\#2,3,4,5,8$. The insets show the spatial distribution of the electric field for the lasing mode $\#1$ (top) and lasing mode $\#6$ (bottom). This pair has the highest SALT interaction coefficient for the D-cavity. The cavities have identical index ($n = 3.5$) and identical surface area, corresponding to a stadium length of $2L = 10~\mu$m.}
\label{fig:BAR}
\end{figure*}

\subsection{Cross-Gain Saturation}
\label{subsection2.25}

The non-universal variation of the number of lasing modes led us to examine the interaction coefficients to see if the results can be explained by cross-saturation. SALT shows us that the quantity which represents modal interactions between lasing modes is the SALT interaction coefficient $\tilde{\chi}_{\mu \nu}$, which we define as 
\begin{equation} 
\tilde{\chi}_{\mu \nu} = \left| \frac{\int_C d\bm{x} \Psi_{\mu} (\bm{x}) \Psi_{\mu} (\bm{x}) |\Psi_{\nu} (\bm{x})|^2}{\int_C d\bm{x} \Psi_{\mu} (\bm{x}) \Psi_{\mu} (\bm{x}) |\Psi_{\mu} (\bm{x})|^2} \right| \, .
\label{eq:chi}
\end{equation} 
Note that mode $\mu$ starts lasing before mode $\nu$ and that the denominator of Eq.~(\ref{eq:chi}) normalizes the interaction coefficient by dividing it by the self-interaction $\chi_{\mu \mu}$ of the mode that started lasing first. Furthermore, we have omitted the Lorentzian factor $\Gamma_\mu$ in $A_{\mu \nu}$, which depends on the relative location of the gain center. 

We calculated the SPA-SALT interaction coefficients for all pairs of modes that start lasing up to the onset of gain clamping. Note that the coefficients describe both the interaction between a pair of modes where both of the modes are lasing, as well as the effect of a lasing mode on second mode before the second one turns on. The values shown for the stadium resonator, Fig.~\ref{fig:BAR}(a), and the D-cavity, Fig.~\ref{fig:BAR}(b), show that there is significant modal interaction and that it has an important effect. For instance, for the stadium, lasing mode $\#4$ does not turn on immediately after the first lasing mode as it would have done in the absence of interactions because of its relatively strong interaction with mode $\#1$ (compared to other modes). Indeed, as shown in the insets, their electric field spatial patterns are very similar, leading to strong cross-saturation of mode $\#4$ by mode $\#1$. Note that these are the two highest-$Q$ modes in the distribution and that they are both strongly scarred by the double diamond periodic orbit. This demonstrates that our results on lasing modes and thresholds are consistent with the expectations for spatial-hole burning and cross-saturation effects. For both the stadium and the D-cavity the interactions influence significantly the order in which modes start to lase. 

\begin{figure*}[!htbp]
\begin{center}
\includegraphics[width = 7.0 cm]{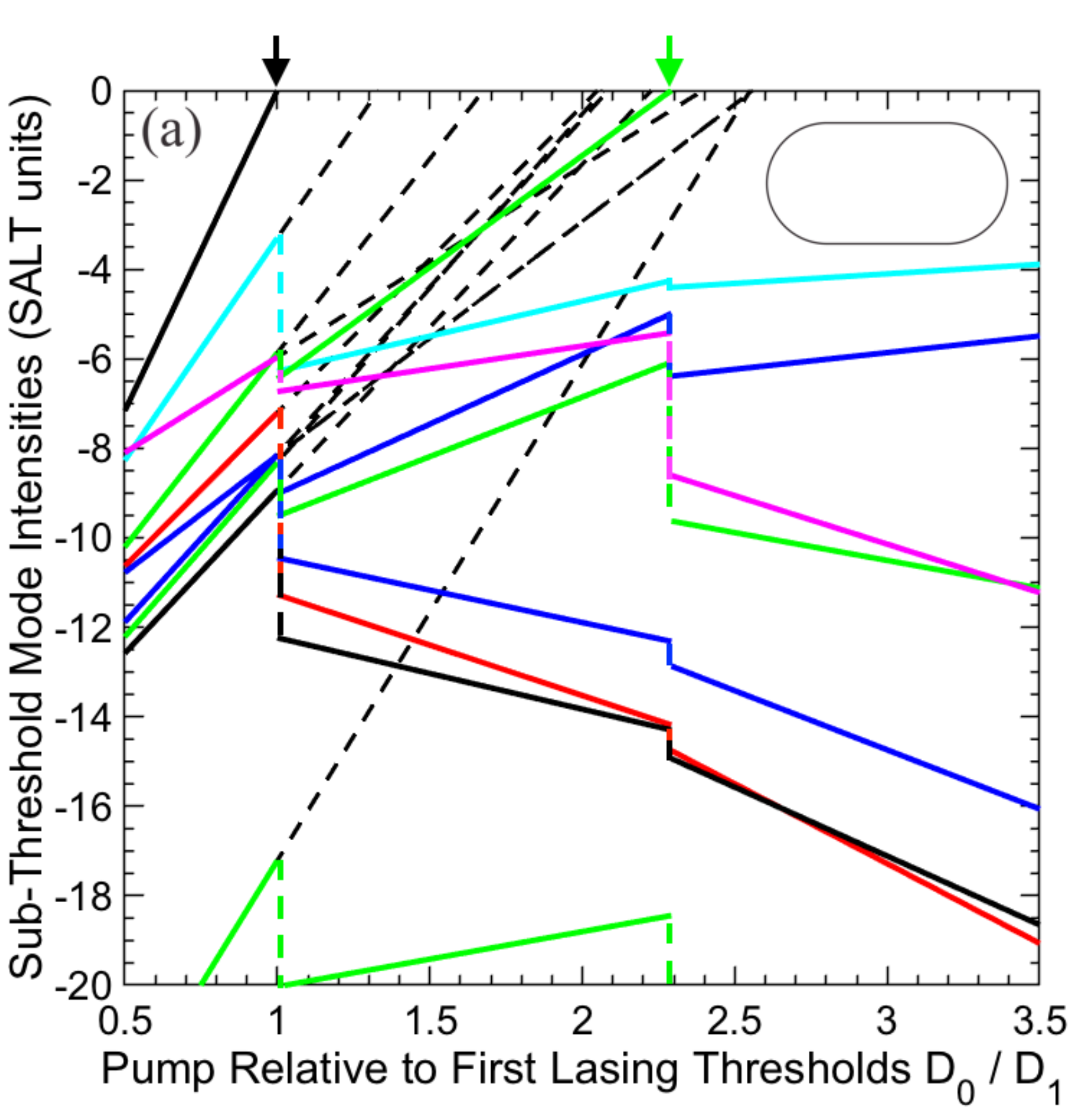}
\includegraphics[width = 7.0 cm]{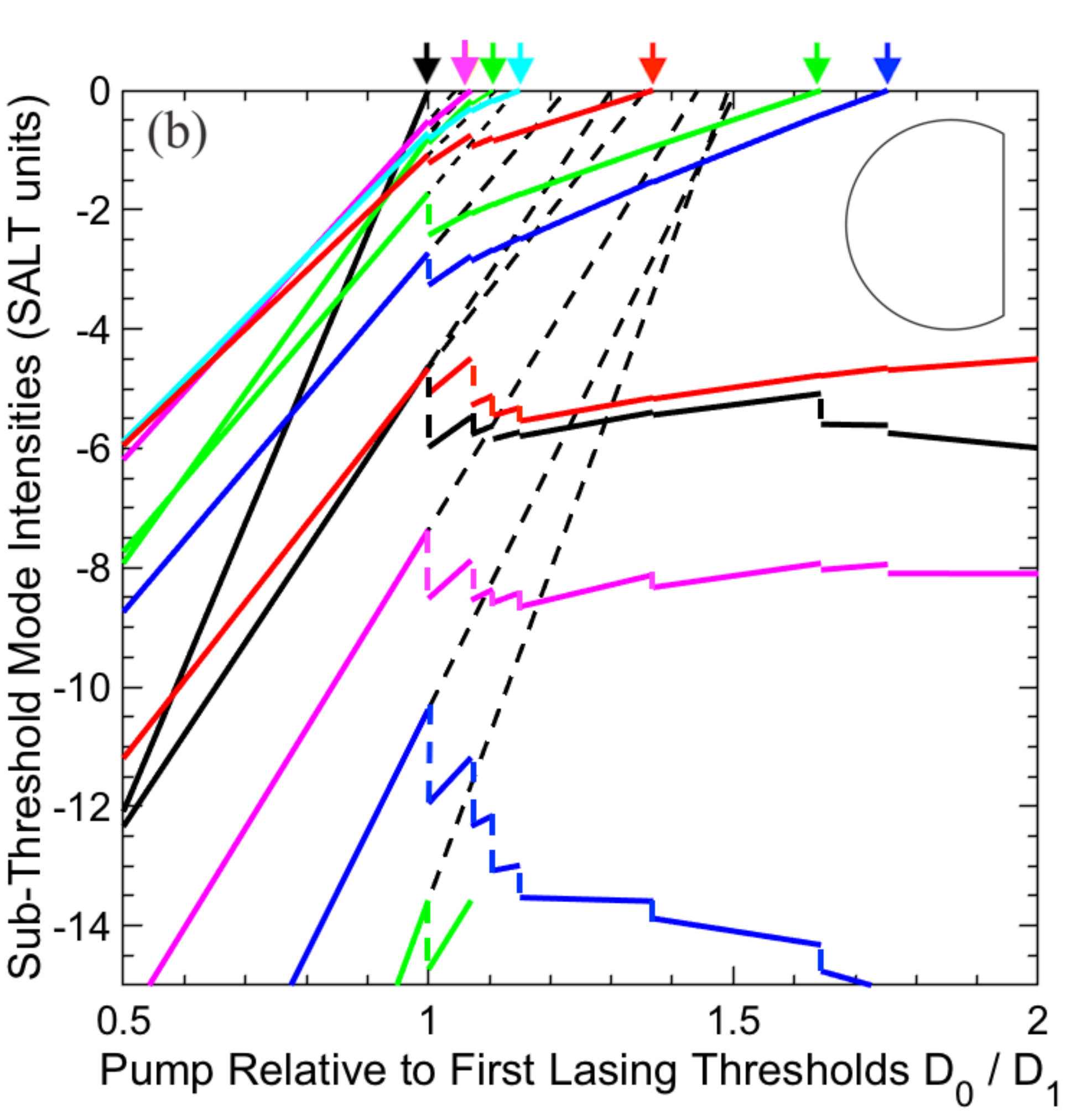}
\end{center}
\caption{Resonance SPA-SALT results for the sub-threshold intensities of the (a)~stadium and the (b)~D-cavity as a function of the pump. The black dashed lines show the intensity variation neglecting modal interactions and intersect the $x$ axis at the non-interacting thresholds. The colored lines show the sub-threshold intensities of the same modes when including the effects of interactions. The resonance SPA-SALT approximations ensure that the sub-threshold intensities are linear between adjacent thresholds, with discontinuities of both their value and slope at the interacting thresholds, which are marked by the colored arrows. The dashed vertical lines serve as a guide for the eye and connect the values of the sub-threshold intensities immediately before and after a new threshold is reached. The gain curve width used in the simulations is $50$~nm. The cavities have identical index ($n = 3.5$) and identical surface area, corresponding to a stadium length of $2L = 10~\mu$m.}
\label{fig:NI}
\end{figure*}

In the case of the D-cavity, the interaction coefficients are more uniform, but there is a strong hole-burning interaction between modes $\#1$ and $\#6$. Even though these modes are not strongly scarred by a single periodic orbit, the insets show a non-uniform spatial pattern as well as a strong similarity of their electric field intensity distributions. In fact there is a weak pseudo-caustic structure in the D-cavity modes which we will discuss elsewhere~\cite{bittner_in_prep}. It appears that the main effect of the absence of high-$Q$ scarred modes in the D-cavity is to allow more modes to lase simultaneously. The D-cavity hence appears to have a $Q$-factor distribution closer to that expected for an ideal wave-chaotic cavity, as described in Ref.~\cite{Schomerus2009}. 

\subsection{Sub-Threshold Intensities}
\label{subsection 2.1}

Not only does spatial-hole burning and gain saturation control the order in which lasing modes turn on, even more importantly it limits their total number. SPA-SALT allows to study the interaction with modes below threshold, which determine when and if a given mode turns on. As noted above, the interacting threshold for a given mode $N$ corresponds to the pump value at which its intensity passes through zero. So just below this threshold, if we expand the size of the $(N-1) \times (N-1)$ matrix $A_{\mu \nu}$, we find that in addition to the positive intensities of the $N -1$ lasing modes, SPA-SALT predicts that mode $N$ has a small negative intensity, approaching zero with positive slope as the pump increases. While this negative intensity is unphysical, its distance below zero, combined with its slope, is a measure of the proximity of the mode from threshold. Hence we introduce here sub-threshold ``intensity" plots to analyze further the effect of modal interactions on modes below threshold. 

As noted above, we compute the sub-threshold negative intensities by enlarging the SPA-SALT matrix $A_{\mu \nu}$ of the lasing modes given in Eq.~(\ref{eq:DAI}), at a given value of the pump, with each of the sub-threshold modes of interest. Between the lasing thresholds these intensities vary continuously and linearly as well as the above threshold modal intensities. As one of the sub-threshold modes reaches zero intensity and turns on, it is added to the physical matrix $A_{\mu \nu}$ of the lasing modes. When this happens, all other sub-threshold modes experience a (typically negative) change of their slope {\it and intensity}. This behavior contrasts with the above threshold, physical modes, which must have continuous intensities as the number of modes increases, although they also have a discontinuity of their slope at the thresholds. 

This threshold effect is calculated separately for each of the non-lasing modes as a way of characterizing the modes that are suppressed from lasing, resulting in a plot of the type shown in Fig.~\ref{fig:NI}. It shows the evolution of the ``negative intensities" of several non-lasing modes as the various lasing modes turn on. The arrows on the top of the figure mark the thresholds of the lasing modes at which the (negative) values of the intensities of the sub-threshold modes have a discrete jump. Note that all of the intensity jumps as well as the slope changes are negative, indicating that each new lasing mode typically reduces the gain for all other modes. Since the evolution of the sub-threshold intensities is linear between thresholds, if the intensity slope of a mode turns negative, this mode will never turn on, no matter how strong the pump becomes (barring the very rare, but not forbidden event that another mode turning on {\it increases} its gain, which is not observed in Fig.~\ref{fig:NI}). When all sub-threshold modes have negative slopes, no further mode can turn on and we have reached the gain-clamping regime. 

Figure~\ref{fig:NI}(a) shows the sub-threshold intensities for the stadium. The black dashed lines show the behavior of the intensities in the absence of interactions; for this case nine lasing modes would turn on within a factor of $3.5$ of the first threshold. However, once the first lasing mode turns on, the modal interactions decrease the values and slopes of the sub-threshold intensities of the other modes so only one more mode turns on. When the second mode turns on, most of the remaining modes obtain negative slopes, so that they can never turn on. Among the modes shown, only the cyan and blue modes in the upper right part of the plot can still turn on, but they get pushed up to thresholds that are many times higher than their non-interacting values which are off the range of the plot. 

Figure~\ref{fig:NI}(b) shows a similar plot of the sub-threshold intensities for the D-cavity. For this case twelve lasing modes would turn on within a factor of $2.0$ of the first threshold in the absence of any interactions. While the threshold of the first mode immediately pushes some of the modes down in intensity, none acquire a negative slope, and the three other modes close to threshold turn on almost immediately, with three more modes turning on within a factor of $2.0$ of the first threshold. Only then are the cumulative interactions sufficiently strong to keep other sub-threshold modes from lasing. This plot clearly shows the effect of the D-cavity having high-$Q$ modes that are closer in their $Q$-factors compared to the stadium with its outliers in the $Q$-factor distribution: the first mode does not have enough intensity to suppress other modes before they turn on. However there are still sufficiently strong interaction effects to keep a number of other modes from lasing and push others to much higher thresholds. Thus, as discussed in Ref.~\cite{Redding2015PNAS}, the total number of lasing modes in steady-state is a function of both the $Q$-factor distribution of the passive cavity modes and the gain competition interactions in the active cavity. 

Whereas the results presented so far were computed for resonators with refractive index $n = 3.5$ (near the experimental value of $n \approx 3.37$), the lasing behavior shows significant dependence on the refractive index, as summarized in Table~\ref{tab:index}. In particular, the few-mode lasing behavior observed for the stadium with $n = 3.5$ is not robust against a change of the refractive index, and for different simulation parameters we see significantly more lasing modes. For the stadium with $n = 2.5$ there are five lasing modes within a factor of $10$ of the first lasing threshold, and eight lasing modes in the same range for $n = 3.0$. Moreover, none of the simulation results show single mode lasing. We conclude that while the number of lasing modes depends on the specific value of the index of refraction of the resonator, multimode lasing is observed for all physically relevant refractive indices. 

\subsection{Dependence on the Size of the Wave-Chaotic Resonators}
\label{subsection2.5}
 
\begin{figure*}[!htbp]
\begin{center}
\includegraphics[width = 15.0 cm]{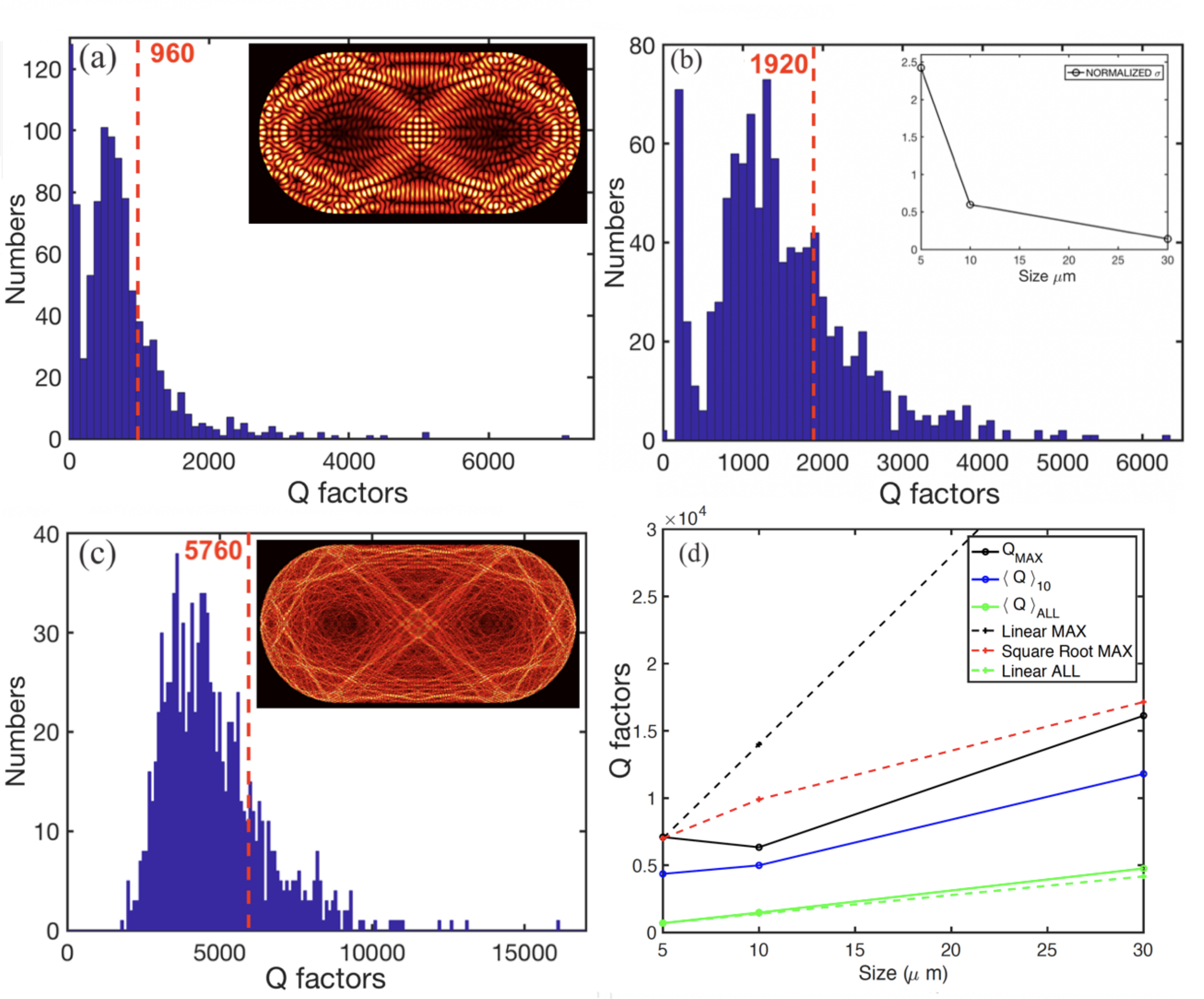}
\end{center}
\caption{Dependence of the $Q$-factor distributions on the size of the stadium resonator. (a)~$2L=10~\mu$m. The inset shows the electric field distribution of the highest-$Q$ mode. (b)~$2L=20~\mu$m. The inset shows the normalized standard deviation of the three $Q$-factor distributions as a function of cavity size. (c)~$ 2L= 60~\mu$m. The inset shows the spatial distribution of the electric field of the highest-$Q$ mode. Notice the qualitative decrease in electric field compared to the highest-$Q$ mode for $2L=10~\mu$m. For panels (a), (b) and (c), the red dashed vertical line marks the $Q$-factor of the long living ray trajectories. (d)~Size dependence of $Q$-factors. Indicated are the highest $Q$-factor, $Q_{MAX}$ (black solid line), the average of the ten highest $Q$-factors, $\left< Q \right>_{10}$ (blue solid line), and the average of all $Q$-factors, $\left<Q\right>_{ALL}$ (green solid line). The linear extrapolations of $Q_{max}$ and $\left<Q\right>_{all}$ from their values at $2L = 10~\mu$m are shown as black and green solid lines, respectively. The red dashed line indicates the extrapolation of $Q_{max}$ according to square-root scaling.}
\label{fig:S13}
\end{figure*}

Our simulation results for a stadium with long axis $2L = 10~\mu$m (area $44.6~\mu \rm{m}^2)$ and a D-cavity of equal area with $n = 3.5$ predict a significant difference between the absolute lasing thresholds of the stadium and the D-cavity lasers due to the $\sim 4.5$ larger $Q$-factor of the highest-$Q$ mode in the stadium compared to that of the D-cavity. This difference, if it could be extrapolated to the larger cavities studied experimentally, would be clearly observable, whereas only a $\sim 1.25$ lower threshold for the stadium was observed. Naively one might expect the $Q$-factors of scarred modes (as well as of all other mode types) to increase linearly with the linear dimensions of the cavity since the $Q$-factor is given by the ratio of energy stored in the resonator, which is proportional to the area, by the energy radiated during one oscillation period, which is proportional to the circumference of the cavity. This simple scaling argument assumes that the general structure of a certain mode type does not depend on the ratio of cavity size and wavelength. It is, however, known that the scarred electric field of a periodic orbit in a given mode field distribution tends to decrease in the semiclassical limit, i.e., when the cavity size becomes much larger than the wavelength \cite{Vergini2015IOP}. Furthermore, the $Q$-factors of scar modes can depend sensitively on interference effects and hence the ratio of cavity size and wavelength \cite{FangAPL2007}. We therefore studied the evolution of the $Q$-factor distributions of stadium and D-cavity resonators with the cavity size to see if the $Q$-factors of high-$Q$ scar modes scale linearly with the cavity size, as well as how the ratio between the highest $Q$-factors of the distributions evolves. 

We studied stadia with linear dimensions two and six times larger than those in Fig.~\ref{fig:Q_DIST}, with the same refractive index $n = 3.5$. The three histograms in Fig.~\ref{fig:S13} show the $Q$-factor distributions for stadia with increasing size, where the long axis $2L$ equals to $10~\mu \rm{m}, 20~\mu$m and $60~\mu$m, respectively. The insets show the electric field distribution of the corresponding highest-$Q$ modes. We observe that most of the high-$Q$ modes have a $Q$-factor increasing with size, but only sub-linearly. In addition, the $Q$-factor distributions narrow at both the high- and low-$Q$ tails, comprised in the stadium by scarred modes (high $Q$-factors) and Bouncing Ball modes (low $Q$-factors). The dashed red vertical lines in the distributions mark the value of the average $Q$-factor of the most long-lived trajectories obtained from ray tracing simulations~\cite{Shinohara2007PRE, Altmann2013RMP, Harayama2015}. As expected, for each distribution this value is higher than the mean of the distribution but lower than a significant number of the high-$Q$ modes. The inset of Fig.~\ref{fig:S13}(b) shows the decrease in the normalized standard deviation of the three distributions, summarizing the narrowing of the distribution with increasing size. Examining in detail the field distributions of the highest-$Q$ modes, [insets of Fig.~\ref{fig:S13}(a) and Fig.~\ref{fig:S13}(c)], we notice a significant reduction in scarred electric field intensity as the linear size of the resonator increases, i.e., the electric field intensity enhancement along the double-diamond orbit is smaller for the larger stadium. 

Figure~\ref{fig:S13}(d) summarizes the evolution of the $Q$-factor variations with size: we plot the highest $Q$-factors, $Q_{MAX}$, the average $Q$-factor of the ten most long-lived modes, $\left< Q \right>_{10}$, as well as the average $Q$-factors of the entire distribution, $\left<Q\right>_{ALL}$, for stadia with long axis equal to $10~\mu$m, $20~\mu$m and $60~\mu$m, respectively. The dashed green line shows that the average $Q$-factor of the entire distribution $\langle Q \rangle_{ALL}$ increases almost linearly with the linear size of the resonator. In contrast, the highest $Q$-factors clearly increase sub-linearly (compare to the black dashed line), and have a scaling behavior closer to a square root law (dashed red line). These results agree with theoretical work that suggests that the effects of scarring decrease as the size of the resonator increases \cite{Vergini2012PRL, Vergini2015IOP} and that the tail of anomalously high $Q$-factors from scarred modes shrinks as well \cite{Novaes2012PRE}. In analogous calculations for the D-cavity comparing the results for sizes $R = 4.2~\mu$m and $R = 8.4~\mu$m (not shown), the average $Q$-factor $\left<Q\right>_{ALL}$ scales almost linearly with the resonator size as well. However, the highest-$Q$ modes are not scarred as strongly as in the case of the stadium, and while their $Q$-factors also increase sub-linearly, there is a smaller difference between the highest-$Q$ modes and the rest of the distribution. We thus conclude that the non-universal effects of scars from short periodic orbits become less important for larger cavities, leading to a similar lasing thresholds for stadium and D-cavity as observed in the experimental data. 

\subsection{Effects of Surface Roughness}
\label{subsection3.1}

\begin{figure*}[!htbp]
\begin{center}
\includegraphics[width = 5.4 cm]{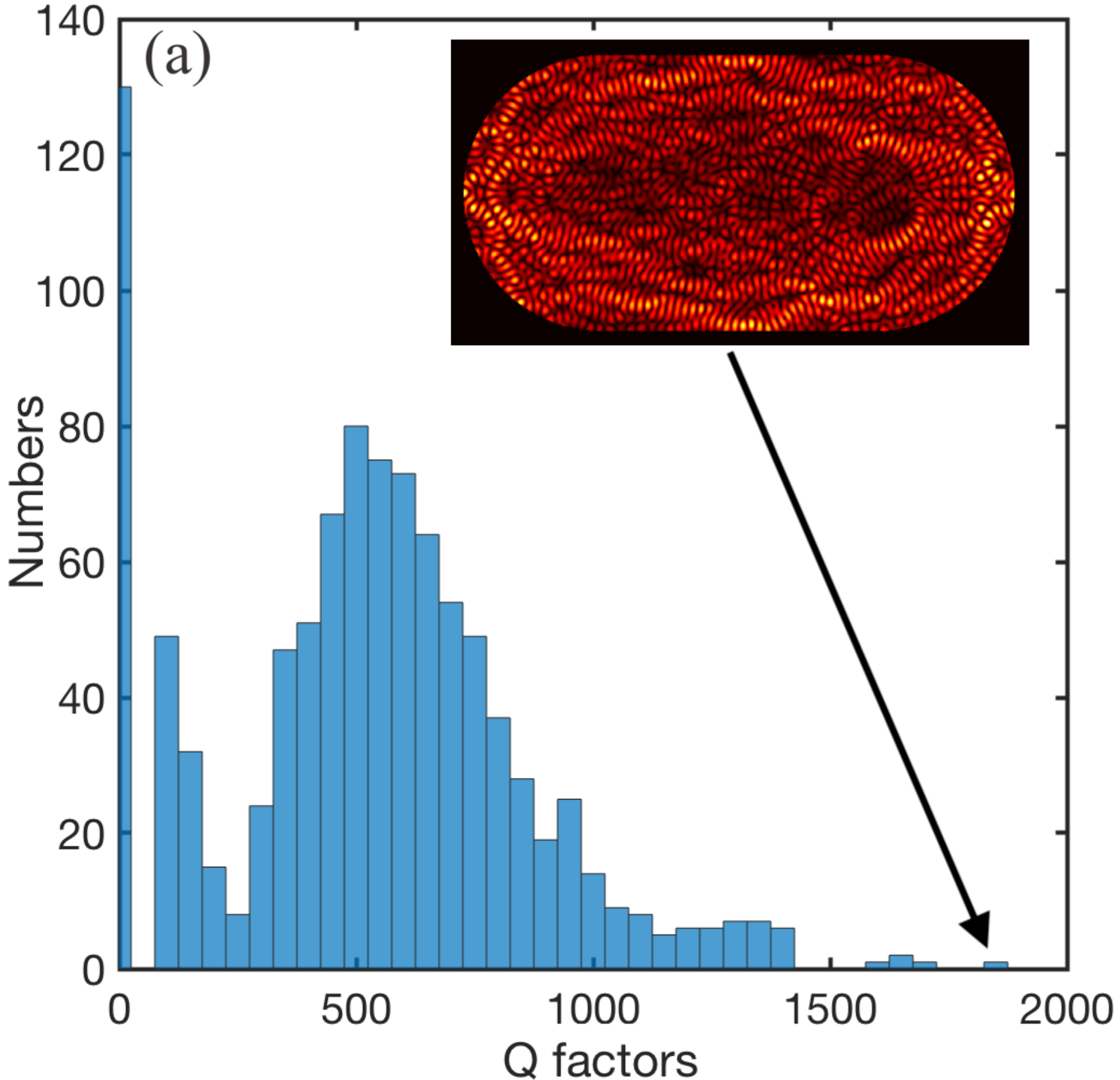}
\includegraphics[width = 5.4 cm]{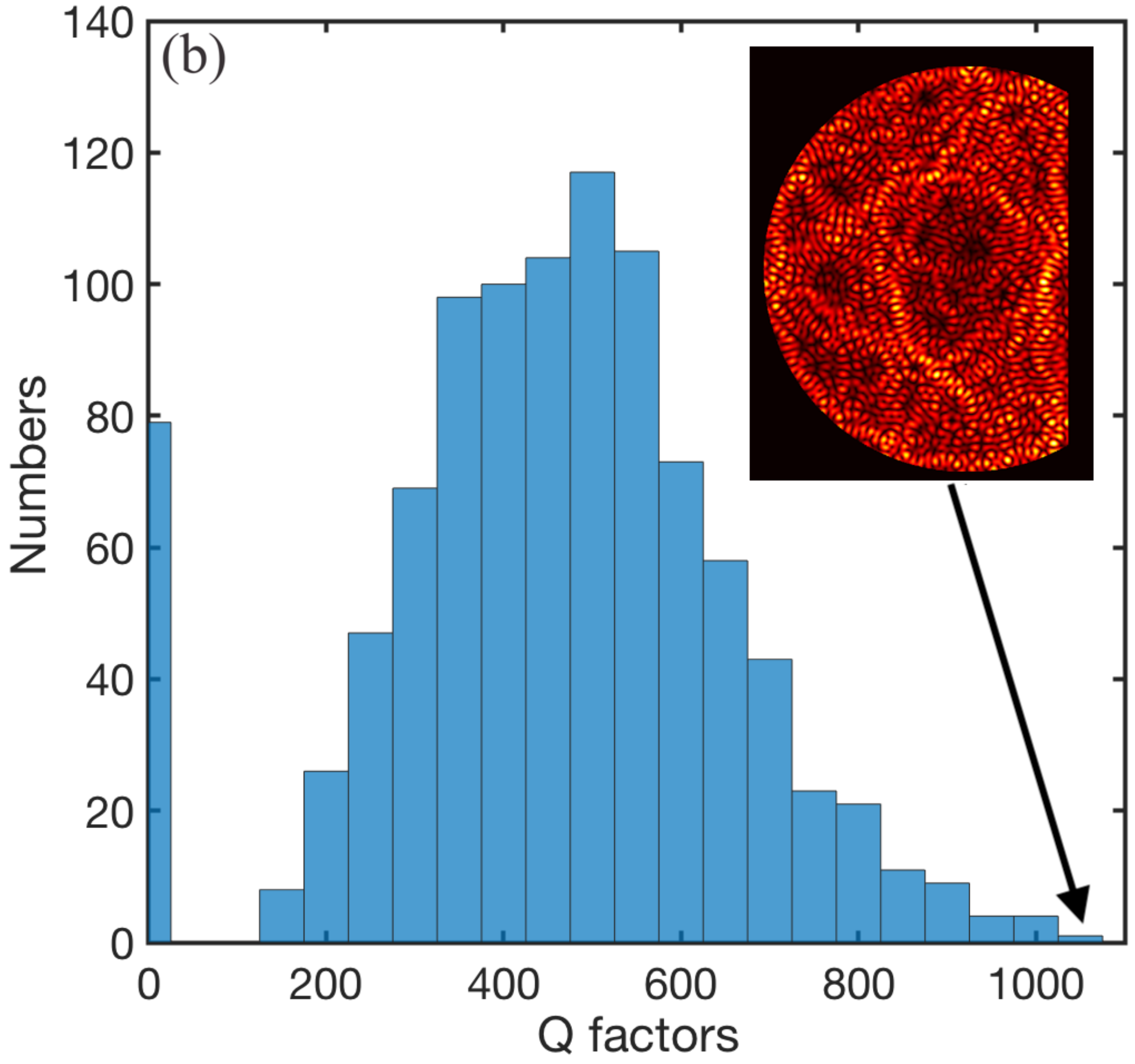}
\includegraphics[width = 5.8 cm ]{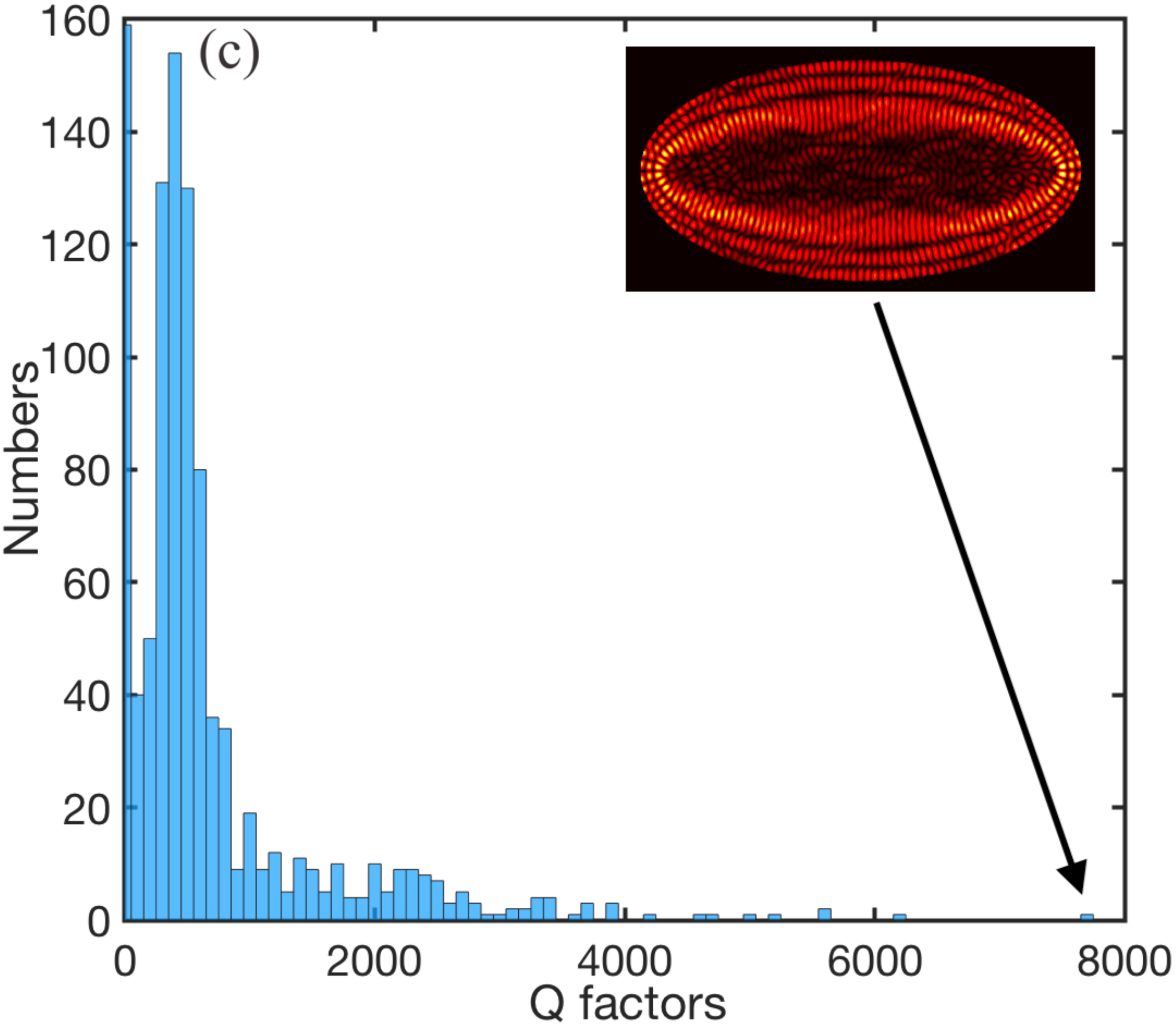}
\end{center}
\caption{$Q$-factor distributions for the (a)~stadium, (b)~D-cavity and (c)~ellipse resonators with surface roughness. Each distribution contains $1000$ resonances centered at $\lambda = 1.0~\mu$m. The random boundary roughness in each of the simulations is drawn from the same distribution, and the standard deviation of the boundary in the normal direction is $\sigma = 30~\rm{nm}$. The insets show the spatial patterns of the modes with the highest $Q$-factors for each distribution. The scarring effect observed for the highest-$Q$ modes of cavities with smooth boundaries do not persist for this level of surface roughness. The cavities have identical index ($n = 3.5$) and identical surface area, corresponding to a stadium length of $2L = 10~\mu$m.}
\label{fig:QSR}
\end{figure*}

Another effect that can result in more similar lasing thresholds for cavities of different shape is surface roughness. Scattering at the rough boundary can affect the field distributions and $Q$-factors in particular of scarred modes. To include this effect in our study, we performed additional simulations in which we randomly modified the smooth boundary of the 2D geometries for three resonator geometries: the stadium, the D-cavity and an elliptical cavity, which has integrable ray dynamics. We introduced random variations in the direction normal to the ideal boundary, with a variation over length scales from $0.3~\mu$m to $10~\mu$m (see Ref.~\cite{liew2014active} for the explicit definition of the deformed boundary). The surface roughness is quantified by the variance of the deviation in normal direction, $r$, from the ideal boundary position, $r_0$. In our simulations the standard deviation of the boundary deformation is $\sigma = \frac{E[(r - r_0)^2]}{r_0}$. Based on our estimate of the surface roughness visible in SEM images of the experimental devices, we used $\sigma = 30~\rm{nm}$. The effects of perturbing the resonator boundary in this way are presented in Fig.~\ref{fig:QSR}, which shows the $Q$-factor distributions for the stadium, D-cavity and ellipse with surface roughness. We notice that the $Q$-factor distribution for the stadium with roughness no longer features high-$Q$ outliers as we suspected. The $Q$-factor distributions of the stadium and D-cavity resonators are much more similar with surface roughness than for a smooth boundary, especially concerning the high-$Q$ tail of the distributions. Still, the average $Q$-factor for the stadium is still higher that of the D-cavity. 

The field distributions of the high-$Q$ modes are also affected as shown in the insets of Fig.~\ref{fig:QSR}, and they deviate significantly from those for smooth boundaries. Therefore the SALT interaction coefficients and thus mode competition are affected as well by the roughness. Table \ref{tab:tablm} summarizes the number of lasing modes within a factor of $10.0$ of the first lasing threshold and the evolution of the $Q$-factors as a function of cavity size for stadium, D-cavity and ellipse. When including surface roughness, doubling the size leads to a roughly two-fold increase of the number of lasing modes, i.e., a linear increase with the cavity size. Note that the simulated ellipse has identical refraction index ($n = 3.5$) and surface area ($A = 44.6~\mu$m) as the other two resonators, and the aspect ratio $b/a = 2$. 

Reference~\cite{Sunada2016PRL}, in which single-mode lasing for stadium-shaped semiconductor microlasers was found, also reported experiments and simulations for elliptical cavities. The ellipse has integrable ray dynamics, and the whispering gallery modes of a dielectric ellipse resonator can be labeled with radial and azimuthal quantum numbers.
The experiments in \cite{Sunada2016PRL} found multimode lasing for elliptical microlasers in contrast to the single-mode lasing of the stadium lasers. It is challenging to model these experiments since an ideal ellipse resonator has extremely high-$Q$ whispering gallery modes, which would lead to many orders of magnitude lower thresholds than observed experimentally. We attribute this to effects such as surface roughness, inaccurate shape fabrication, and intrinsic absorption that reduce the $Q$-factors of actual resonators. We simulated elliptical microlasers with surface roughness equal to that used for the wave-chaotic resonators. We found that the rough ellipse cavities always showed multimode lasing, with a number of modes comparable to, but smaller than for the wave-chaotic cavities, as shown in Table~\ref{tab:tablm}.  The highest $Q$-factor for the rough ellipse was four times higher than that for the rough stadium, and the mode showed a more regular field distribution despite the surface roughness. This indicates that an integrable cavity shape, like the ellipse or the circle, has resonance properties which are distinct from the wave-chaotic cavities, even in the presence of substantial surface roughness. 

\begin{table}[!htbp]
\begin{tabular}{lcccc}
\hline \hline
Resonator & $Q_{MAX}$ \quad  & \quad $\left<Q\right>_{10}$ \quad &  \quad LM  \\
\hline
Stadium SR $2L = 10~\mu$m & $ 1833  $  & $ 1524 $ & $ 9 $   \\
D-cavity SR $2R = 8.4~\mu$m & $ 1034  $  & $ 969 $   & $ 11 $  \\
Ellipse SR $2b = 10.6~\mu$m   & $ 7662 $   & $ 5282 $  & $ 6 $  \\
Stadium SR $2L = 20~\mu$m  & $ 2852  $  & $ 2724 $ & $ 15 $  \\
D-cavity SR $2R = 16.8~\mu$m  & $ 2083  $  & $ 1895 $ & $ 22 $  \\
Ellipse SR $2b = 21.2~\mu$m    & $ 16495 $ & $ 13800 $ & $ 14 $  \\
\hline \hline
\end{tabular}
\caption{Highest $Q$-factor ($Q_{MAX}$) and the average of the $10$ highest $Q$-factors ($\left<Q\right>_{10}$), as well as the number of Lasing Modes (LM) within a factor of $10.0$ of the first lasing threshold for the stadium, D-cavity and ellipse with surface roughness. The refractive index is $n = 3.5$ and the geometrical parameters are chosen such that the the different geometries in the top half of the table have the same area, as do the ones in the bottom half of the table.}
\label{tab:tablm}
\end{table}

\section{Discussion and Conclusion}
\label{conclusion}

We present an experimental and theoretical study concerning the question if wave-chaotic semiconductor microlasers in cw operation generally exhibit multimode lasing, as was found for experiments with pulsed pumping in Ref.~\cite{Redding2015PNAS} and steady-state numerical simulations in Ref.~\cite{Cerjan2016Controlling}, or single-mode lasing as found in experiments with cw pumping and theoretical studies in Refs.~\cite{2013SunadaPRA, Sunada2016PRL, Harayama2017PhotonRes}. 

Our new experimental results, presented in the first part of this paper, can be summarized as follows. We find multimode lasing for both stadium and D-cavity lasers in qualitative agreement with our earlier results on the D-cavity \cite{Redding2015PNAS}. Time-resolved measurements for long pump pulses show fewer lasing peaks at a given time compared to integration over a whole pulse, but we do not observe a consistent reduction of the number of active lasing modes towards a single mode over the course of longer pulses as was found in Ref.~\cite{2013SunadaPRA}. The time-resolved measurements showed that the spectra were stable over time scales very long compared to other time scales in the system, so we believe our results are representative of the steady-state lasing properties. The devices studied in the current experiments have lasing thresholds comparable to those studied in Ref.~\cite{Sunada2016PRL}, thus it is unlikely that a different degree of surface roughness is responsible for our observation of multimode lasing in contrast to single-mode lasing in the experiments of Refs.~\cite{2013SunadaPRA, Sunada2016PRL}.  

The new theoretical and simulation results presented have the following implications in our view. Non-universal effects, such as lasing on scarred modes, become already weak in wave-chaotic cavities with dimensions well below the size of the experimental cavities, and cannot explain single-mode lasing in wave-chaotic microlasers. SALT theory, which takes into account gain competition and spatial hole burning very accurately, never predicts single-mode lasing for wave-chaotic resonators pumped well above threshold. Hence gain saturation and spatial overlap of modes alone do not explain the experimental results found in Ref.~\cite{Sunada2016PRL}. Hence the remaining theoretical uncertainties are in the dynamics of the gain medium, which is neglected in SALT. It is well known that for nearly degenerate lasing modes, a frequency locking effect can occur above threshold due to the population dynamics, reducing the number of lasing modes~\cite{Harayama2003PRL}. FDTD simulations of wave-chaotic lasers by the Harayama and coworkers~\cite{Harayama2017PhotonRes, Harayama2017SELF} appear to show such merging of modes as the pump is increased, but these effects are found at unphysically high relative pumps, and at large values of the population relaxation rate $\gamma_\parallel$. If there is a regime where single-mode lasing is favored for wave-chaotic lasers, it must be due to subtle dynamical effects such as these, which go beyond standard spatial hole-burning and gain competition.  We currently lack the theoretical and computational tools to answer definitively this question and we did not explicitly consider certain effects that might impact the dynamics, such as carrier diffussion. Our experimental results indicate that if there is such a regime, it is not realized in all wave-chaotic semiconductor lasers, and needs to be better characterized experimentally. 

\acknowledgements

This work was supported by AFOSR Award No. FA9550-16-1-0416.

The authors thank T. Harayama, T. Fukushima, S. Sunada, S. Shinohara, M. Guy and K. Kim. 

\newpage

\bibliography{MULTIMODE_BIB}

\end{document}